\documentclass[12pt]{article}
\usepackage{amsmath,latexsym,epsf}
\usepackage{epsfig,amsfonts,cite}
\usepackage{amssymb}
\usepackage{graphicx}

\def\tmod{|\tau|}

\long\def\comment#1{}

\def\phizero{\phi(0)}
\def\psizero{\psi(0)}
\def\hzero{h(0)}
\def\phil{\phi(l)}
\def\psil{\psi(l)}
\def\hl{h(l)}
\def\phizero{\phi_0}
\def\psizero{\psi_0}
\def\hzero{h_0}
\def\phil{\phi}
\def\psil{\psi}
\def\hl{h}
\def\gzero{g_0}
\def\gl{g}
\def\yg{y_g}
\def\yTg{y_{\phi g}}
\def\yT{y_\phi}
\def\yH{y_h}
\def\ypsideux{y_{\psi^2}}
\def\ypsitrois{y_{\psi^3}}
\def\yTpsi{y_{\phi\psi}}
\def\yHpsi{y_{h \psi}}
\def\nuH{\nu_h}
\def\muH{\mu_h}
\def\yTn{y_{\phi_n}}
\def\yTun{y_{\phi_1}}
\def\yTdeux{y_{\phi_2}}
\def\yTtrois{y_{\phi_3}}
\def\yHn{y_{h_n}}
\def\yHun{y_{h_1}}
\def\yHdeux{y_{h_2}}
\def\yHtrois{y_{h_3}}

\def\Eq#1{Eq.~(\ref{#1})}
\def\Eqs#1#2{Eqs.~(\ref{#1}-\ref{#2})}

\begin{document}

\pagestyle{myheadings}
\markright{version of \today,  Nancy}

\title{Numerical revision of the universal amplitude ratios for the two-dimensional 4-state Potts model}
\date{\today}
\author{
  {Lev N. Shchur$^{*,**}$, Bertrand Berche$^{**}$ and Paolo Butera$^{***}$}\\
  {\small\it $^*$ Landau Institute for Theoretical Physics, }  \\[-0.2cm]
  {\small\it Russian Academy of Sciences,}  \\[-0.2cm]
  {\small\it Chernogolovka 142432, Russia}  \\[-0.2cm]
  \\
  {\small\it $^{**}$ Laboratoire de Physique des Mat\'eriaux, }  \\[-0.2cm]
  {\small\it Universit\'e Henri Poincar\'e, Nancy 1}     \\[-0.2cm]
  {\small\it BP 239, F-54506 Vand\oe uvre les Nancy Cedex, France} \\[-0.2cm]
  \\
  {\small\it $^{***}$ Istituto Nazionale di Fisica Nucleare, }  \\[-0.2cm]
  {\small\it  Sezione di Milano-Bicocca, }  \\[-0.2cm]
  {\small\it Piazza delle Scienze 3, 20126, Milano,   Italia}\\[-0.2cm]
  \\
  {\small {\tt lev@landau.ac.ru, }}\\[-0.2cm]
  {\small {\tt berche@lpm.u-nancy.fr, }}\\[-0.2cm]
  {\small {\tt paolo.butera@mib.infn.it}}\\[-0.2cm]
  {\protect\makebox[5in]{\quad}}
  \\
}
\vspace{0.5cm}

\newcommand{\be}{\begin{equation}}
\newcommand{\ee}{\end{equation}}
\newcommand{\bey}{\begin{eqnarray}}
\newcommand{\eey}{\end{eqnarray}}
\newcommand{\<}{\langle}
\renewcommand{\>}{\rangle}

\maketitle
\thispagestyle{empty}

\vspace{0.2cm}

\begin{abstract}

Monte Carlo (MC) simulations and series expansion (SE) data for the
energy, specific heat, magnetization and susceptibility of the
ferromagnetic 4-state
Potts model on the square lattice are analyzed in a vicinity
of the critical point in order to estimate universal combinations of
critical amplitudes. The quality of the fits is improved using
predictions of the renormalization group (RG) approach and of
conformal invariance, and restricting the data within an appropriate
temperature window.

The RG predictions on the cancelation of the logarithmic corrections
in the universal amplitude ratios are tested.
A direct calculation of the effective ratio of the energy amplitudes using
duality relations explicitly demonstrates this cancelation of
logarithms, thus supporting the predictions of  RG.

We emphasize the role of corrections {\em and} of background terms
on the determnination of the amplitudes.
The ratios
of the critical amplitudes of the susceptibilities obtained in our
analysis differ significantly from those predicted theoretically
and supported by
earlier SE and MC analysis. This disagreement might signal that the
two-kink approximation used in the analytical estimates is not
sufficient to describe with fair accuracy the amplitudes of the
4-state model.

\end{abstract}

%
%

\section{Introduction}
\label{sec-intro}

In a first paper~\cite{I} (hereafter denoted as I), we studied the
universal combinations of critical amplitudes of the 3-state Potts
model. The present paper is devoted to a similar  analysis in the
4-state case, which is much more involved  due to the presence of
logarithmic corrections  strongly influencing
the critical behavior.

We analyze numerical data obtained in Monte Carlo (MC) simulations using
the Wolff~\cite{Wolff89} single-cluster algorithm and also the series
expansion (SE) data available in the literature. In the following, we
refer to the data type as MC and SE data, respectively. For comparison
with our own results, we  shall also reconsider the data obtained in MC
simulations using the Swendsen-Wang cluster algorithm~\cite{Swendsen-Wang}
by Caselle, Tateo, and Vinci~\cite{CaselleTateoVinci99} and indicated
as CTV data.

Our motivation in using different data sets is to achieve a better
control of the critical behavior, since one may expect, for the
three sources, some differences in the critical region due to the
different interplay of the finite size effects. In addition, one can
apply different techniques to the data analysis: the fits in the case of
the MC data
and the approximant technique in the case of the
 SE data. The consistency of the
final results will increase our confidence.
We care so much because the presence of logarithmic corrections makes
the numerical determination of the critical behavior of the 4-state
Potts model a rather delicate task.

To be even safer, two different approaches are used, which were
successfully applied also to the $q=3$ state Potts model in I. First
we estimate the critical amplitudes, which are   then used   to
compute universal ratios. Second, besides the direct determination
of the amplitudes themselves, we estimate {\em ratios} of critical
amplitudes, constructing effective ratio functions, and computing
their limiting values at the critical point. This provides a direct
estimate of universal ratios. Analyzing the renormalization group
equations, we have shown in an Appendix (see
also~\cite{SBB_EPL,SBB_CSCMP}) that, in the absence of any regular
background term, the logarithmic corrections cancel in the effective
ratio functions.

The Hamiltonian of the ferromagnetic  Potts model~\cite{Potts52} reads as
\begin{equation}
    H = - \sum_{\langle ij \rangle}\delta_{s_i s_j}\; ,
    \label{Ham}
\end{equation}
where $s_i$ is a ``spin'' variable taking integer values between $0$
and $q-1$, and the sum is restricted to the nearest neighbor sites
$\langle ij \rangle$ on a  lattice of $N$ sites with periodic
boundary conditions. The partition function $Z$ is defined by
\begin{equation}
    Z =  \sum_{conf} {\rm e}^{-\beta H}.
    \label{Z-part}
\end{equation}
 with $\beta=1/k_BT$~\footnote{According to the
usual terminology, the inverse temperature and the critical
exponent of the magnetization  are denoted by the same symbol $\beta$, since
there is no risk of confusion in this context.},
and $k_B$ the Boltzmann constant (fixed to unity).
On the square lattice in zero magnetic field, the model is self-dual.
Denoting by $\beta^*$ the dual of the inverse temperature $\beta$,
the  duality relation
\begin{equation}
    \left( e^\beta - 1 \right) \left( e^{\beta^*} - 1 \right)=q
    \label{d-t}
\end{equation}
 determines  the critical value of the inverse temperature~\cite{Potts52}
$\beta_c=\ln (1+\sqrt{q})\approx 1.09861$.
Dual reduced temperatures $\tau$ and $\tau^*$ can be defined by
\begin{equation}
\beta=\beta_c(1-\tau)\;\;\; {\rm and}\;\;\; \beta^*=\beta_c(1+\tau^*).
\label{dual-tt}
\end{equation}
Close to the critical point, $\tau$ and
$\tau^*$ coincide through first order, since $\tau^*=\tau+\frac{\ln
(1+\sqrt q)}{\sqrt q}\tau^2+O(\tau^3)$.

The critical amplitudes and the critical exponents describe the
singular behavior of the thermodynamic quantities close to the
critical point.  For example the magnetization $M$, the (reduced)
susceptibility $\chi$ and the specific heat $C$ of a spin system in
zero external field\footnote{In this paper we only deal with the
physical properties in zero magnetic field.}  behave as
\footnote{Note that for simplicity  we have dropped for the moment
the multiplicative logarithmic corrections and allowed only for
additive corrections.}
\begin{eqnarray}
    M(\tau) &=& B (-\tau)^\beta \left(1+{\rm corr.\; terms}\right),\ \tau <0 \label{m-crit}\\
    \chi(\tau)_\pm &=& \Gamma_\pm \tmod^{-\gamma}\left(1+{\rm corr.\; terms}\right), \label{x-crit}\\
     C(\tau)_\pm &=& \frac{A_\pm}{\alpha}\tmod^{-\alpha}\left(1+{\rm corr.\; terms}\right),
    \label{c-crit}
\end{eqnarray}
 Here $\tau$ is the reduced temperature $\tau=(T-T_c)/T$ and the
labels $\pm$ refer to the high-temperature (HT) and low-temperature
(LT) sides of the critical temperature $T_c$. In addition to the
mentioned observables, for the Potts models with $q>2$ a transverse
susceptibility can be defined in the low-temperature phase\footnote{In
the following we shall use the notations $\Gamma_L$ or $\Gamma_T$ for
the longitudinal or transverse susceptibility amplitudes in the
low-temperature phase. When still used, $\Gamma_-$  is identified with
$\Gamma_L$.}
\begin{equation}
    \chi_T(\tau) = \Gamma_T (-\tau)^{-\gamma}\left(1+{\rm corr.\; terms}\right).
    \label{xt-crit}
\end{equation}

The critical exponents are known exactly for the 2D Potts
model~\cite{denNijs79,Pearson80,Nienhuis84,Dotsenko84,DotsenkoFateev84}:
\begin{equation}
    x_\epsilon=\frac{1+y}{2-y},\quad    x_\sigma=\frac{1-y^2}{4(2-y)},
    \label{x-Tm}
\end{equation}
where $y$ is related to the number of states $q$
of the Potts variable by
\begin{equation}
    \cos\frac{\pi y}{2}=\frac12 \sqrt{q}.
    \label{y-from-q}
\end{equation}
 The standard exponents follow from
 $x_\epsilon=(1-\alpha)/\nu$
and  $x_\sigma=\beta/\nu$.
The central charge of the corresponding conformal field theory is
also simply expressed~\cite{Dotsenko84,DotsenkoFateev84} in terms of
$y$
\begin{equation}
    c=1-\frac{3y^2}{2-y}.
    \label{central-c}
\end{equation}

Analytical estimates of critical amplitude ratios for the $q$-state
Potts models with $q=1$, $2$, $3$, and $4$ were recently obtained by
Delfino and Cardy~\cite{DelfinoCardy98}. They used the exact 2D
scattering field theory of Chim and
Zamolodchikov~\cite{ChimZamolodchikov92} and estimated the ratios
using a two-kink approximation for $1<q\le 3$. For $3<q\le4$, they
considered both the two-kink approximation and the contribution from
the bound state. For $q=4$ this approximation leads  to the value
$c=0.985$ for the central charge, to be compared to the exact value
$c=1$.  Using this approximate value, one can calculate the scaling
dimensions from (\ref{central-c}) and (\ref{x-Tm}) obtaining the
values $x_\sigma=0.117$ and $x_\epsilon=0.577$, to be compared with
the exact values $1/8$ and $1/2$ respectively. So, the deviation
from the exact values becomes as large as 6 and 15 per cent,
emphasizing the difficulty of the $q=4$ case. In the 3-state case
the situation is much better (see I).

Let us recall that the existence of  logarithmic corrections to
scaling  in the 4-state Potts model was pointed out in  the
pioneering works of Cardy, Nauenberg and
Scalapino~\cite{CardyNauenbergScalapino80,NauenbergScalapino80},
where a set of non-linear RG equations was solved. Their discussion
was  recently extended by Salas and Sokal~\cite{SalasSokal97}.

The universal susceptibility amplitude ratio $\Gamma_+/\Gamma_L$
was calculated by Delfino and Cardy in~\cite{DelfinoCardy98} for
both the 3-state and 4-state Potts models.
 Later Delfino, et
al.~\cite{DelfinoBarkemaCardy00} estimated analytically also the
ratio of the transverse to the longitudinal susceptibility
amplitudes $\Gamma_T/\Gamma_L$. The values obtained in the 4-state
case are
\begin{eqnarray}
&\Gamma_+/\Gamma_L=4.013, & \Gamma_T/\Gamma_L=0.129 .
    \label{res-theor-q4}
\end{eqnarray}
In this latter paper~\cite{DelfinoBarkemaCardy00}, the results of
MC  simulations were also reported, but they
were considered inconclusive by the authors.
  More recently   Delfino
and Grinza report compatible values in the case of the
Ashkin-Teller model, using the same technique at the
same level of approximation~\cite{DelfinoGrinza04}.

Another contribution to the study of the amplitude ratios in the 2D
4-state Potts model was reported by Caselle, et
al~\cite{CaselleTateoVinci99}. These authors presented a MC
determination of various amplitudes.  In particular, their estimate
of the susceptibility amplitude ratio $\Gamma_+/\Gamma_L=3.14(70)$
is   in reasonable agreement with
the theoretical estimate of Delfino and Cardy, in
spite of a somewhat controversial~\cite{DelfinoBarkemaCardy00} use
of the logarithmic corrections in the fitting procedure.

Enting and Guttmann~\cite{EntingGuttmann03} also analyzed SE data
for the 4-state Potts model and found
\begin{eqnarray}
&\Gamma_+/\Gamma_L=3.5(4), & \Gamma_T/\Gamma_L=0.11(4),
    \label{res-EG-q4}
\end{eqnarray}
  results which are compatible   with the predictions
of~\cite{DelfinoCardy98} and~\cite{DelfinoBarkemaCardy00}. Their
series analysis does not rely on differential approximants, but, in
the hope to achieve better control of the log-corrections of the
$q=4$ case, they address directly the asymptotic behavior of the
series coefficients.

In the present paper we present more accurate MC data
supplemented by a reanalysis of the extended series derived
by Enting and Guttmann~\cite{EntingGuttmann03}.
 We address the following question: Is it possible to
estimate the influence of the logarithmic corrections on the fit
procedure? Is it possible to devise some procedure in which the
role of the logarithmic corrections is properly taken into account?

  In the rest of the paper,   we
shall be concerned with the following universal combinations of
critical amplitudes
\begin{equation}
    \frac{A_+}{A_-}, \;\; \frac{\Gamma_+}{\Gamma_L}, \;\;
    \frac{\Gamma_T}{\Gamma_L}, \;\; R_C^+=\frac{A_+\Gamma_+}{B^2},
    \;\; R_C^-=\frac{A_-\Gamma_L}{B^2}
    \label{univ-rat}
\end{equation}
where the last two are  a consequence of the scaling
relation\footnote{We refer the reader to the review
Ref.~\cite{PrivmanHohenbergAharony91} for a detailed
discussion of the universality of the critical amplitudes ratios.}
$\alpha=2-2\beta-\gamma$. To the various critical amplitudes of
interest, $A_\pm$, $\Gamma_\pm$,\dots, we have associated
appropriately defined ``effective amplitudes'', namely
temperature-dependent quantities $A_\pm(\tau)$,
$\Gamma_\pm(\tau),\ldots$, which take as limiting values, when $\tau
\rightarrow 0$, the critical amplitudes $A_\pm$, $\Gamma_\pm, \dots$.
In order to avoid any risk of confusion between the critical amplitude and
the   corresponding
effective temperature-dependent amplitude, reference to these
temperature-dependent quantities is always made with their explicit
$\tau-$dependence.

Also in this paper, we make use of the duality relation in order to
improve the estimates of the  ratios between effective amplitudes
measured at {\em dual} temperatures.  In the case of the 4-state
Potts model, this procedure {\em would even eliminate all
logarithmic corrections} from the fit in absence of background
contributions, which unfortunately do exist for most quantities!  We
again use the duality relation to estimate the correction-to-scaling
amplitudes in the behavior of the specific heat and of the
susceptibility. For this purpose, we compute ratios also on the
duality line, e.g. the
 susceptibility effective-amplitude ratio
$\Gamma_+(\tau)/\Gamma_L(\tau^*) =\chi_+(\beta)/\chi_L(\beta^*)$ as
the ratio of $\chi_+(\beta)$, the high-temperature susceptibility at
inverse temperature $\beta$, and of $\chi_L(\beta^*)$, the
low-temperature susceptibility at the dual inverse temperature
$\beta^*$. Furthermore, we show analytically that the leading
logarithmic corrections cancel on the duality line for the ratio of
the specific-heat amplitudes as extracted from the energy at dual
temperatures.

As a final result of our analysis\footnote{The figures given here
are an average between the MC and the SE
determinations~\cite{SBB_EPL}.}, we propose estimates of the
susceptibility critical-amplitude ratios $\Gamma_+/\Gamma_L=6.49(44)$
 and $\Gamma_T/\Gamma_L=0.154(12)$ which
are significantly different both from the predictions eq.(\ref{res-theor-q4})
of~\cite{DelfinoCardy98,DelfinoBarkemaCardy00}, and from the
numerical estimates of~\cite{CaselleTateoVinci99,EntingGuttmann03}.
The deviation from the  numerical estimates of other authors
 might be explained
by the complicated logarithmic corrections used to fit the data
in~\cite{CaselleTateoVinci99,EntingGuttmann03}. The difference from
the theoretical predictions might be  due to the limited accuracy of
the approximation scheme used  for $q>3$.

In conclusion, obtaining an accurate (say, within a few per cent) and
generally accepted approximation of the critical amplitude
combinations for the four-state Potts model still remains an open
issue both theoretically and numerically.

%
%
\section{Computational procedures}
\label{sec-model}

%
%
\subsection{Monte Carlo simulations}
\label{sec-MC}

We use the single-cluster Wolff algorithm~\cite{Wolff89} for studying
square lattices of linear size $L$ with periodic boundary
conditions. Starting from an ordered state, we let the system
equilibrate in $10^5$ steps measured by the number of flipped Wolff
clusters. The averages are computed over $10^6$---$10^7$ steps. The
random numbers are produced by an exclusive-XOR combination of two
shift-register generators with the taps (9689,471) and (4423,1393),
which are known~\cite{Shchur99} to be safe for the Wolff algorithm.

The order parameter of a microstate
${\tt M}({\tt t})$ is evaluated during the simulations as
\begin{equation}
    {\tt M}=\frac{qN_m/N-1}{q-1},
    \label{Order-Potts}
\end{equation}
where $N_m$ is the number of sites $i$ with $s_i=m$ at the time
$\tt t$ of the simulation~\cite{Binder81}, and $m\in [0,1,...,(q-1)]$
is the spin value  of the majority of the spins. $N=L^2$ is the
total number of spins. The thermal average is denoted $M=\<{\tt M}\>$.

Thus, the longitudinal susceptibility in the low-temperature phase is
measured by the fluctuation of the majority spin orientation
\begin{equation}
    k_BT\; \chi_L=\frac{1}{N} (\langle N_m^2\rangle-\langle N_m\rangle^2)
    \label{susc-LT}
\end{equation}
and the transverse susceptibility is defined
 in terms of the fluctuations of the minority of the spins
\begin{equation}
    k_BT\; \chi_T=\frac{1}{(q-1)N}\sum_{\mu\ne m}
    (\langle N_\mu^2\rangle-\langle N_\mu\rangle^2),
    \label{susc-T}
\end{equation}
while in the high-temperature phase $\chi_+$ is given by the fluctuations
in all $q$ states,
\begin{equation}
    k_BT\; \chi_+=\frac{1}{qN}\sum_{\mu=0}^{q-1}
    (\langle N_\mu^2\rangle - \langle N_\mu\rangle^2),
    \label{susc-HT}
\end{equation}
where $N_\mu$ is the number of sites with the spin in the state $\mu$.
Properly allowing for the finite-size effects, this definition of the
susceptibilities is, in both phases, completely consistent with the
available SE data~\cite{ShchurButeraBerche02}.

The internal energy density of a microstate is calculated as
\begin{equation}
    {\tt E}=-\frac{1}{N} \sum_{\langle ij \rangle}\delta_{s_i s_j}\,
    \label{energy}
\end{equation}
its ensemble average is  denoted by $E=\<{\tt E}\>$
and the reduced specific heat per spin measures the energy fluctuations,
\begin{equation}
      (k_BT)^2  \; C=-\frac{\partial  E }{\partial\beta}
    =\langle {\tt E}^2 \rangle - \langle {\tt E} \rangle^2.
    \label{heat}
\end{equation}

We have simulated the model on square lattices with linear sizes
$L=20$, $40$, $60$, $80$, $100$ and $200$.  In each case, we have
measured the physical quantities within a range of reduced temperatures
called the ``critical window'' and defined as follows. Assuming a
proportionality factor of order 1 in the definition of the
correlation length, the relation $L < \xi \propto \tmod^{-\nu}$
yields the value of the reduced temperature at which the correlation
length becomes comparable with the system size $L$ and thus below
which the finite-size effects are not negligible.
This value defines the lower end of the critical
window and avoids finite size effects which would  make our analysis
more complex. The upper limit of the critical window is fixed for
convenience at $\tau=0.20-0.25$.

%
%
%
\subsection{Series expansions}
\label{sec-SE}

Our MC study of the critical amplitudes will be supplemented by an
analysis of the high-temperature and low-temperature expansions for
$q=4$, recently extended through remarkably high orders by Briggs,
Enting and Guttmann ~\cite{BriggsEntingGuttmann94,EntingGuttmann03}.
In terms of these series, we can compute the effective critical
amplitudes for the susceptibilities, the specific heat and the
magnetization and extrapolate them by the standard resummation
techniques, namely simple Pad\'e approximants (PA) and differential
approximants (DA), properly biased with the exactly known critical
temperatures and critical exponents.

The LT expansions are expressed in terms of the variable
$z=\exp(-\beta)$. In the $q=4$ case, the expansion of the energy extends
through $z^{43}$. For the longitudinal susceptibility the expansion
extends through $z^{59}$, and for the transverse susceptibility
through $z^{47}$. In the case of the magnetization, the expansion
extends through $z^{43}$. The HT expansions are computed in terms of
the variable $v=(1-z)/(1+(q-1)z)$. They extend up to $v^{43}$ in the
case of the energy and up to $v^{24}$ for the susceptibility.

It is useful to point out that, for convenience, in
Ref.~\cite{EntingGuttmann03} the product of the susceptibility by
the factor $q^2/(q-1)$, rather than the susceptibility itself, is
tabulated at HT, because it has integer expansion coefficients.
 For the same reason, at LT  the magnetization times $q/(q-1)$ is
tabulated.  Therefore the appropriate normalization should be
restored in order that the series yield amplitudes consistent with
the MC results.

As a general remark on our series analysis, we may point out that
the accuracy of the amplitude estimates  given in Ref. I for the
$q=3$ case is good due to the relatively harmless nature of the
power-like corrections to scaling, while in the $q=4$ case  the
mentioned resummation methods cannot reproduce the  expected
logarithmic corrections to scaling and therefore the extrapolations
to the critical point are more uncertain. In this case we have
tested  also a somewhat unconventional use of DA's: in computing the
effective amplitudes, we only retain DA estimates outside some small
vicinity of the critical point, where they appear to be stable and
reliable. Finally, we perform the extrapolations by fitting these
data  to an asymptotic form which includes logarithmic corrections.
We shall add further comments on the specific analyses in the next
sections.
%

\section{Critical amplitudes of 4-state Potts model}
\label{sec-4PM}
\subsection{Expected temperature-dependence of the observables}

In the case of the 4-state Potts model, we have $y=0$  from
(\ref{y-from-q}) and the second thermal
exponent~\cite{Nienhuis82,Dotsenko84,DotsenkoFateev84}
$\yTdeux=-4y/3(1-y)$ vanishes. Accordingly, the leading
power-behavior of the specific heat (and of other physical
quantities) is modified~\cite{CardyNauenbergScalapino80} by a
logarithmic factor
 \begin{equation}
    {C}(\tau)=\frac{A_\pm}{\alpha} \tmod^{-\alpha}(-\ln\tmod)^{-1}
    {\cal X}_{corr}(-\ln\tmod)+{\cal Y}_{bt}(\tmod).
    \label{C4}
\end{equation}
The exponent $\alpha$ takes the value $2/3$ and ${\cal
X}_{corr}(-\ln\tmod)$ contains correction terms with powers of
$\tmod$, $-\ln\tmod$ and $\ln(-\ln\tmod)$. It may also contain
non-integer power corrections due to the higher thermal
exponents~\cite{denNijs79,Pearson80,Nienhuis84,Dotsenko84,DotsenkoFateev84}
$\yTn$ or to other irrelevant fields, as well as power corrections
due to the identity field.
  ${\cal Y}_{bt}(\tmod)$ contains all regular contributions and is
referred to as the background term.

Extending the pioneering work of Cardy, Nauenberg and
Scalapino~\cite{CardyNauenbergScalapino80,NauenbergScalapino80},
Salas and Sokal~\cite{SalasSokal97} obtained a set of non-linear RG
equations. In the appendix (see also
Refs.~\cite{SBB_EPL,SBB_CSCMP}), we derive from this set of
equations a closed expression for the leading logarithmic
corrections, which is more suitable to describe the temperature
range accessible in a numerical study, than the asymptotic form
given by Salas and Sokal, \be {C}(\tau)=\frac{A_\pm}{\alpha}
\tmod^{-2/3}(-\ln\tmod)^{-1}\left[ 1-\frac
32\frac{\ln(-\ln\tmod)}{-\ln\tmod} +O\left(\frac
1{\ln\tmod}\right)\right],\label{eq-C_SS} \ee which is the first term
of a slowly convergent expansion of ${\cal X}_{corr}(-\ln\tmod)$ in
logs. We have observed that the following expansion (see Appendix)
is better behaved in the temperature window near the critical point
accessible by MC and SEs
\bey {C}(\tau)&=&\frac{A_\pm}{\alpha}
|\tau|^{-2/3}{\cal G}^{-1}(-\ln|\tau|),
        \label{eq-C_us}\\
    {\cal G}(-\ln|\tau|)&=&(-\ln|\tau|)\times{\cal E}(-\ln|\tau|)\times {\cal F}(-\ln|\tau|),
    \label{Eq-Dfn_H}\\
    {\cal E}(-\ln|\tau|)&=&
    \left(1+\frac 34\frac{\ln(-\ln|\tau|)}{-\ln|\tau|}\right)
        \nonumber\\
    &&
    \ \times\left(1-\frac 34\frac{\ln(-\ln|\tau|)}{-\ln|\tau|}\right)^{-1}
    \left(1+\frac 34\frac{1}{(-\ln|\tau|)}\right),
        \label{Eq-Dfn_G}\\
    {\cal F}(-\ln|\tau|)&\simeq&
    \left(1+\frac{C_1}{-\ln|\tau|}
    +\frac{C_2\ln(-\ln|\tau|)}{(-\ln|\tau|)^2}\right)^{-1}.\label{eq-f59}
\eey The function $\cal E(-\ln\tmod)$ contains the exact form of the
leading terms with universal coefficients  predicted by RG. The
remaining part is made of log terms, whose
coefficients involve the non-universal dilution field $\psizero$.
The multiplicative function ${\cal F}(-\ln|\tau|)$ mimics, in a
given temperature range close to the critical point, the higher-order
terms of this non-universal part, which is a slowly convergent series
in powers of logs starting with an $O\left(\frac 1{\ln\tmod}\right)$
term. The values of $C_1$ and $C_2$ in (\ref{eq-f59})  should not be
considered as ``real'' amplitudes of correction-to-scaling terms, but
only as ``effective'' parameters. A ``zeroth-order'' analysis may be
performed taking $C_1=C_2=0$, i.e. ${\cal F}(-\ln\tmod)=1.$ A more
refined estimate follows from the analysis of the magnetization, as
explained in the section~\ref{subsecMq4} and in the Appendix.  The
absence of a constant background term is a simplifying feature in the
analysis of the magnetization\footnote{The effective amplitudes are
constructed by dividing the corresponding physical quantity by the
leading terms (main power dependence with known logarithmic
corrections), and, therefore the background terms, if present, will be
divided by the logarithmic terms. These terms may seriously
complicate the analysis of the
limiting behavior of the effective amplitudes.}. The
difference in behavior between the two types of expressions
(\ref{eq-C_SS}) and (\ref{eq-C_us}) for the specific heat is
illustrated by the fact that, for example, at a typical value of
temperature numerically accessible without finite-size effects,
$\tau=0.10$, we have
 $1-\frac32\frac{\ln(-\ln\tmod)}{-\ln\tmod}\simeq 0.457 $, while
$\left(1-\frac 34\frac{\ln(-\ln|\tau|)}{-\ln|\tau|}\right) \times
\left(1+\frac 34\frac{\ln(-\ln|\tau|)}{-\ln|\tau|}\right)^{-1}
\simeq 0.786$ and the two functions are not simply proportional to
each other in the typical range $\tau =0.02 - 0.20$.
  Fitting the numerical data with one or the other choice
may thus spoil the outcome for the leading amplitude.

Similar expressions of the logarithmic corrections are obtained for
the other physical quantities in \Eqs{ap-eq100}{ap-eq101} of the
Appendix.

Further corrections to scaling   may also be present
in \Eq{eq-C_us}. They
 are discussed in the
Appendix and may be of the form   $a_{2/3}|\tau|^{2/3}$,
$a_{h_3}|\tau|^{\Delta_{h_3}}$ or
$a_{\phi_3}|\tau|^{\Delta_{\phi_3}}$, as well as powers of these
terms, where $\Delta_{h_3}=3/4$ and $\Delta_{\phi_3}=5/3$. Pure
power corrections and background terms may also be needed. Here we
also stress that the inclusion of a leading correction in
$a|\tau|^{2/3} $ and of analytic terms seems to be necessary
according to the papers by Joyce~\cite{Joyce75a,Joyce75b}, where the
magnetization of a model, expected to belong to the 4-state Potts
model universality class, is shown to have an expression of the form
\be M(-\tmod)=\tmod^{1/12}(f_0(\tau)+\tmod^{2/3}f_1(\tau)
+\tmod^{4/3}f_2(\tau))\label{eq-M_Joyce} \ee with $f_i(\tau)$
analytic functions when $\tau\to 0^-$. The correction exponents are
obtained from the table of the conformal scaling dimensions by
Dotsenko and Fateev~\cite{Dotsenko84,DotsenkoFateev84}, but not all
of them are necessarily present. However, at least the presence of
the exponents $2/3$, and possibly of $4/3$, seems to be needed in
order to account for the numerical results. Caselle et
al.~\cite{CaselleTateoVinci99} also considered an
$a_{2/3}|\tau|^{2/3} $ term to fit the magnetization. The parameter
$a_{h_3}$ is a priori possibly needed only for magnetic quantities,
while the corrections in $a_{\phi_3}$ will systematically be dropped
in our fits, since they are sub-sub-dominant.

 In conclusion, the most
general expression that we will consider is the following: \bey
    {\rm Obs.}(\pm|\tau|)
        &\simeq&{\rm Ampl.}\times
    \tmod^{\blacktriangleleft}\times
    {\cal G}^{\bigstar}(-\ln\tmod)\times(1+{\rm corr.\ terms})+\nonumber\\
    &\ &\qquad\qquad+{\rm\  backgr.\ terms},
        \label{eq-Obs_us}\\
    {\rm corr.\ terms}&=&a_{2/3} \tmod^{2/3}+b_\pm\tmod+a_{4/3}\tmod^{4/3}+\dots,\\
    {\rm backgr.\ terms}&=&D_0+D_1\tmod+\dots
\eey where ${\cal G}(-\ln\tmod)$ is defined by
Eqs.~(\ref{Eq-Dfn_H}-\ref{eq-f59}), while  ${\blacktriangleleft}$ and
$\bigstar$ stand for exponents which depend on the observable considered.
They are all given in the Appendix.
%

\subsection{The magnetization amplitude\label{subsecMq4}}
The amplitude $B$  of the magnetization  is defined by the
asymptotic behavior  (see the appendix for details)
\begin{eqnarray}
    M(-|\tau|)&=&B|\tau|^{1/12}(-\ln|\tau|)^{-1/8}
    \left[
    \left(1+\frac 34\frac{\ln(-\ln|\tau|)}{(-\ln|\tau|)}\right)
    \left(1-\frac 34\frac{\ln(-\ln|\tau|)}{(-\ln|\tau|)}\right)^{-1}
        \right.\nonumber\\
    &&\left.
    \ \qquad\times
    \left(1+\frac 34\frac{1}{(-\ln|\tau|)}\right)
    {\cal F}(-\ln|\tau|)
        \right]^{-1/8}
            (1+a\tmod^{2/3}+b\tmod+\dots).\label{Eq-FullM}
\end{eqnarray}
We can extract an effective function ${\cal F}_{eff}(-\ln|\tau|)$
which mimics the real one ${\cal F}(-\ln|\tau|)$ in the convenient
temperature range  $|\tau|\simeq 0.01-0.10$. This is done by
plotting  an effective magnetization amplitude
\begin{eqnarray}
    B_{eff}(-|\tau|)&=&M(-|\tau|)|\tau|^{-1/12}(-\ln|\tau|)^{1/8}
    \left[
    \left(1+\frac 34\frac{\ln(-\ln|\tau|)}{(-\ln|\tau|)}\right)
        \right.\nonumber\\
    &&\left.
    \qquad\qquad\qquad
    \times\left(1-\frac 34\frac{\ln(-\ln|\tau|)}{(-\ln|\tau|)}\right)^{-1}
    \left(1+\frac 34\frac{1}{(-\ln|\tau|)}\right)
        \right]^{1/8}
\label{b-4-eff}
\end{eqnarray}
which is then fitted to the expression
\be
    B_{eff}(-|\tau|)=B\left(1+\frac {C_1}{-\ln\tmod}
+\frac{C_2\ln(-\ln|\tau|)}{(-\ln|\tau|)^2}\right)^{1/8}\times (1 +a
\tmod^{2/3}+b\tmod+\dots) \label{eq-Beff} \ee in which we have also
included corrections to scaling.  As we have
already noticed, the coefficients appearing in the function  ${\cal F}$ are
effective parameters adapted to the temperature window considered, therefore
the values of $C_1$ and $C_2$ have no special meaning.

\begin{figure}[ht]
  \centering
  \begin{minipage}{\textwidth}
  \epsfig{file=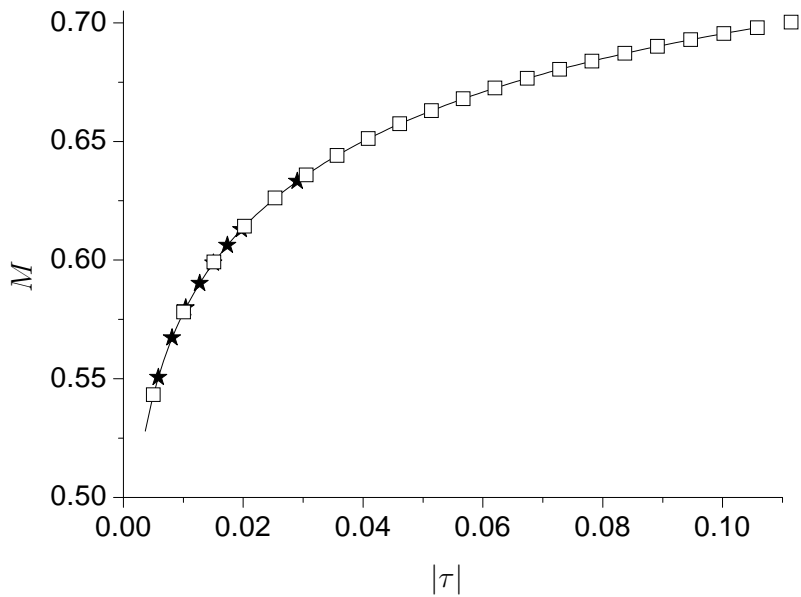,width=0.60\textwidth}
  \end{minipage}
  \vspace{-4cm}
\caption{\small The magnetization $M$ in the critical window region. Our MC
data are represented by boxes, the MC data from
Ref.~\cite{CaselleTateoVinci99} by stars
and the SE data  by a solid line.}
    \label{magn-all}
\end{figure}

In order to analyze the numerical data and to extract
the different coefficients, one needs very accurate data.
As an illustration, in Figure~\ref{magn-all} we compare
MC and SE data, and MC data from
Caselle et al.~\cite{CaselleTateoVinci99}.

\begin{figure}[ht]
  \centering
  \begin{minipage}{\textwidth}
  \epsfig{file=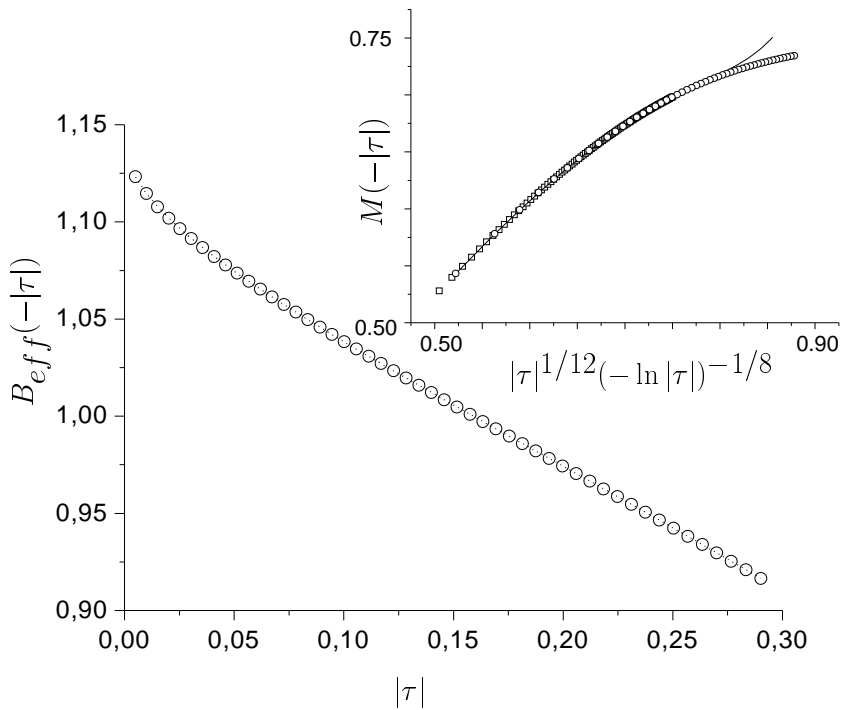,width=0.60\textwidth}
  \end{minipage}
  \vspace{-3cm}
\caption{\small The effective amplitude of the magnetization
    $M$. Insert: The magnetization $M$ as function of
    $\tmod^{1/12}(-\ln\tmod)^{-1/8}$. (Our MC data are represented by
  open circles, the SE data by boxes
    and the fit by a solid line).}
    \label{FigBeff}
\end{figure}

The behavior of $B_{eff}(-\tmod)$ is shown in Fig.~\ref{FigBeff}. In
table~\ref{table_parameters_M}, we present a selection of our fits of
MC data to \Eq{eq-Beff}. The first column of the table indicates the
different choices of the function ${\cal F}(-\ln\tmod)$. For each line
in the tables, several fits have been tried, varying the number of
points in the interval $\tmod\in[0.005,0.25]$ (total number of points
50) and calculating the $\chi^2/d.o.f.$ for each fit. A reasonable
balance has to be found between the distance of the points from the
critical temperature and their number. It appeared that limiting the
fit window to 20 data points, i.e. to the interval $[0.005,0.1]$,
gives the best confidence level. This choice is quite satisfactory,
since it corresponds to a close vicinity of the critical point. The
criterion that we adopted in order to select a most convincing fit
among all possible fits is the stability of the correction-to-scaling
amplitudes $a$ and $b$ in equation~(\ref{Eq-FullM}) when the
temperature window is varied, typically in the range
$\tmod\in[0,0.06]$ to $\tmod\in[0,0.30]$. As long as these numbers
have not converged to given values, we cannot pretend that it is
meaningful to include such corrections in the analysis. We have to
mention that stability of the correction amplitudes is never reached
if we stop the correction terms in equation~(\ref{Eq-FullM}) at the
leading log, or even at the three higher log terms (correction
function ${\cal E}(-\ln\tmod)$).  Then even the leading amplitude $B$
is questionable. The effective function ${\cal F}_{eff}(-\ln|\tau|)$
is essential to reach convergence of all but the two effective
coefficients $C_1$ and $C_2$   which still strongly depend on the
temperature window.

For a given type of fit defined in the first column,
Tables~\ref{table_parameters_M} and \ref{table_parameters_M-SE} show
the parameters of the fit minimizing $\chi^2$ and the minimization is
performed by varying the width and position of the temperature window.

Once our choice is made for $C_1$ and $C_2$, the results of the fit
for the amplitude $B$ show a remarkable stability. The entry called
fit \# 0 corresponds to a ``zeroth-order'' fit in which the function
${\cal F}(-\ln\tmod)$ is taken equal to unity. It is presented for
comparison, in order to emphasize the improvement occurring in the
following lines.  Fit \#1 keeps all correction coefficients, while the
lines which follow are obtained  decreasing the number of fit
parameters. We favor fits \# 2 and \#3, where the coefficient $b$ (the
amplitude of the linear term $\tmod$) is fixed to zero, since from the
first fit   (\#1)   we see that leaving $b$ free leads to a value close to zero
 and we consider more reliable a fit
with fewer parameters. Fit \# 4 corresponds to another extreme choice including
 no irrelevant correction at all. It shows that the effective function
${\cal F}(-\ln\tmod)$ is more important than the irrelevant
corrections to scaling in order to achieve a stable amplitude $B$
(comparable to the outcomes of fits \# 1 to \# 3).

\begin{table}[ht]
\caption{\small Various fits of our MC data for
 the magnetization effective amplitude
 $M(-\tmod)\tmod^{-1/12}(-\ln\tmod)^{1/8} {\cal E}^{1/8}(-\ln\tmod)$
 to the expression~\Eq{eq-Beff}.  The stars in the first column
 indicate our favorite fits (the reasons for this choice are given in
 the text).}  \center\scriptsize
\begin{tabular}{llllll}
\hline\noalign{\vskip-1pt}\hline\noalign{\vskip2pt} Fit \# &
Amplitude & \multispan{4}{\hfil Correction terms \hfil}
\\
\multispan{1}{\vrule height0pt width0mm} & \multispan{1}{\vrule
height0.5pt width17mm} & \multispan{4}{\vrule height0.5pt width73mm}
\\
\noalign{\vskip2pt}
        & $B$  & $\propto\frac 1{-\ln|\tau|} $
        & $\propto\frac {\ln(-\ln\tmod)}{(-\ln|\tau|)^2} $
    & $\quad\propto|\tau|^{2/3}$ & $\quad\propto|\tau|$
       \\
\noalign{\vskip2pt} \hline
MC \# 0 & $1.1355(4)$ & $\phantom- -$ & $\phantom- -$ & $-0.41(1)$&$\phantom- 0.41(21)$ \\
MC \# 1 & $1.1566(5)$ & $-0.740(2)$ & $-0.630(51)$ & $-0.172(4)$&$-0.018(6)$ \\
MC \# $2^*$ & $1.1570(1)$ & $-0.757(1)$ & $-0.522(11)$ & $-0.191(2)$ & $\phantom- -$ \\
MC \# $3^*$ & $1.1559(12)$ & $-0.88(5)$ & $\phantom- -$ & $-0.21(1)$ & $\phantom- -$  \\
MC \# 4 & $1.1593(1)$ & $-0.25(2)$ & $-3.04(5)$ & $\phantom- -$ & $\phantom- -$ \\
\hline\noalign{\vskip-1pt}\hline\noalign{\vskip2pt}
\end{tabular}
\label{table_parameters_M}
\end{table}
\begin{table}[ht]
\caption{\small Same as table~\ref{table_parameters_M} for the
magnetization data obtained by the SE method. A star in the first column
indicates our favorite fit with the coefficients $C_1$ and $C_2$
(now fixed and shown in bold face in this table and in the forthcoming
tables) obtained from the
fits of the MC data reported in table~\ref{table_parameters_M}.}
\center\scriptsize
\begin{tabular}{llllll}
\hline\noalign{\vskip-1pt}\hline\noalign{\vskip2pt} Fit \# &
Amplitude & \multispan{4}{\hfil Correction terms \hfil}
\\
\multispan{1}{\vrule height0pt width0mm} & \multispan{1}{\vrule
height0.5pt width17mm} & \multispan{4}{\vrule height0.5pt width73mm}
\\
\noalign{\vskip2pt}
        & $B$  & $\propto\frac 1{-\ln|\tau|} $
        & $\propto\frac {\ln(-\ln\tmod)}{(-\ln|\tau|)^2} $
    & $\quad\propto|\tau|^{2/3}$ & $\quad\propto|\tau|$
       \\
\noalign{\vskip2pt} \hline
SE \# 0 & $1.1364(4)$ & $\phantom- -$ & $\phantom- -$ & $-0.435(5)$&$\phantom- 0.097(10)$ \\
SE \# 1 & $1.1597(1)$ & $-0.637(5)$ & $-1.417(28)$ & $-0.115(2)$&$\phantom- 0.013(2)$ \\
SE \# $2$ & $1.1589(1)$ & $-0.660(5)$ & $-1.246(25)$ & $-0.126(2)$ & $\phantom- -$ \\
SE \# $2^*$ & $1.1575(1)$ & $-{\bf 0.757}$ & ${\bf -0.522}$ & $-0.194(1)$ & $\phantom- -$ \\
SE \# $3$ & $1.1583(8)$ & $-0.981(34)$ & $\phantom- -$ & $-0.185(10)$ & $\phantom - -$  \\
SE \# $3^*$ & $1.1575(1)$ & $ {\bf -0.88}$ & $\phantom- -$ & $-0.225(1)$ & $\phantom - -$  \\
SE \# 4 & $1.1573(1)$ & $\phantom- 0.164(7)$ & $-4.106(18)$ & $\phantom- -$ & $\phantom- -$ \\
\hline\noalign{\vskip-1pt}\hline\noalign{\vskip2pt}
\end{tabular}
\label{table_parameters_M-SE}
\end{table}

\begin{table}[ht] \caption{\small Same as table~\ref{table_parameters_M}
 but using the MC data of Caselle et
    al.~\cite{CaselleTateoVinci99} for the magnetization.
The fit marked by a star is performed
 using the same coefficient $C_1$ (shown in bold) as in
table~\ref{table_parameters_M}. }
\center\scriptsize
\begin{tabular}{llllll}
\hline\noalign{\vskip-1pt}\hline\noalign{\vskip2pt} Fit \# &
Amplitude & \multispan{4}{\hfil Correction terms \hfil}
\\
\multispan{1}{\vrule height0pt width0mm} & \multispan{1}{\vrule
height0.5pt width17mm} & \multispan{4}{\vrule height0.5pt width82mm}
\\
\noalign{\vskip2pt}
        & $B$  & $\propto\frac 1{-\ln|\tau|} $
        & $\propto\frac {\ln(-\ln\tmod)}{(-\ln|\tau|)^2} $
    & $\quad\propto|\tau|^{2/3}$ & $\quad\propto|\tau|$
       \\
\noalign{\vskip2pt} \hline
CTV \# 0 & $1.1386(2)$ & $\phantom- -$ & $\phantom- -$ & $-0.518(6)$&$\phantom- 0.30(2)$ \\
CTV \# 1 & $1.196\pm1.499$ & $-7.5\pm163$ & $\phantom- 19.6\pm429$ & $-1.09\pm17.3$&$1.2\pm29.2$ \\
CTV \# 2 & $1.148(6)$ & $\phantom- 0.45(56)$ & $-3.6\pm1.1$ & $-0.17(4)$ & $\phantom- -$ \\
CTV \# 3 & $1.162(9)$ & $-1.13(38)$ & $\phantom- -$ & $-0.14(12)$ & $\phantom- -$  \\
CTV \# $3^*$ & 1.1561(1) & ${\bf -0.88}$ & $\phantom- -$ & $-0.215(1)$ &$\phantom- -$ \\
CTV \# 4 & $1.153(2)$ & $\phantom- 0.73(31)$ & $-5.5\pm 0.77$ & $\phantom- -$ & $\phantom- -$ \\
\hline\noalign{\vskip-1pt}\hline\noalign{\vskip2pt}
\end{tabular}
\label{table_parameters_MCaselle}
\end{table}

The same procedure is now applied to the SE data. It is a
non-conventional approach to fit SE data, which are usually analyzed
by approximant methods, but the exercise is tempting. We thus apply
exactly the same procedure as in the case of MC data, varying the
fitting interval and comparing the values of the $\chi^2/d.o.f.$.
The best results are collected in table~\ref{table_parameters_M-SE}.
The agreement between the results quoted in the two tables is
amazing. Not only the amplitudes, but also the correction
coefficients are very close to each other. In this table (and in the
forthcoming tables), we also present the results of this analysis of
SE data with $C_1$ and $C_2$ {\em fixed} to their best values
extracted in table~\ref{table_parameters_M} from the MC data. They
are indicated again with a star (SE $\# 2^*$ and SE $\# 3^*$) and in
order to emphasize the fact that $C_1$ and $C_2$ in this case are
not free parameters, they are indicated in bold face. A reason for
this approach is first to insist on the consistency of the results
and second to privilege a fit with less free parameters (and in
particular no free log term). For the magnetization amplitude, a
compromise between  MC and SE data analysis provides our final
estimate

\be B=1.157(1).\ee

As an additional test, we decided to fit also the MC data of Caselle
et al.~\cite{CaselleTateoVinci99}, obtained using the Swendsen-Wang
cluster algorithm, to our functional expression~\Eq{eq-Beff}.
Notice that the various coefficients reported in
table~\ref{table_parameters_MCaselle} (not only the critical
amplitude, but also the correction terms) are very close to those
reported above in table ~\ref{table_parameters_M}. In particular
this fit gives further support to our choice in
favor of fits \# 2 and \# 3.

In the following tables, we will refer to the best two fits by the
labels \# $2^*$ and \# $3^*$ (which means that the coefficients $C_1$
and/or $C_2$ are fixed   to their values
indicated in table~\ref{table_parameters_M})   when fitting other
quantities.
  Note that
the MC data of Caselle et al.~\cite{CaselleTateoVinci99} are not
fitted with the choice \# $2^*$, because they do not extend on a range
of temperatures wide enough to apply the corresponding functional
expression.

%
%
\subsection{The specific heat and the energy density}\label{subsecEq4}

\begin{figure}[ht]
  \centering
  \begin{minipage}{\textwidth}
  \hskip-2cm\epsfig{file=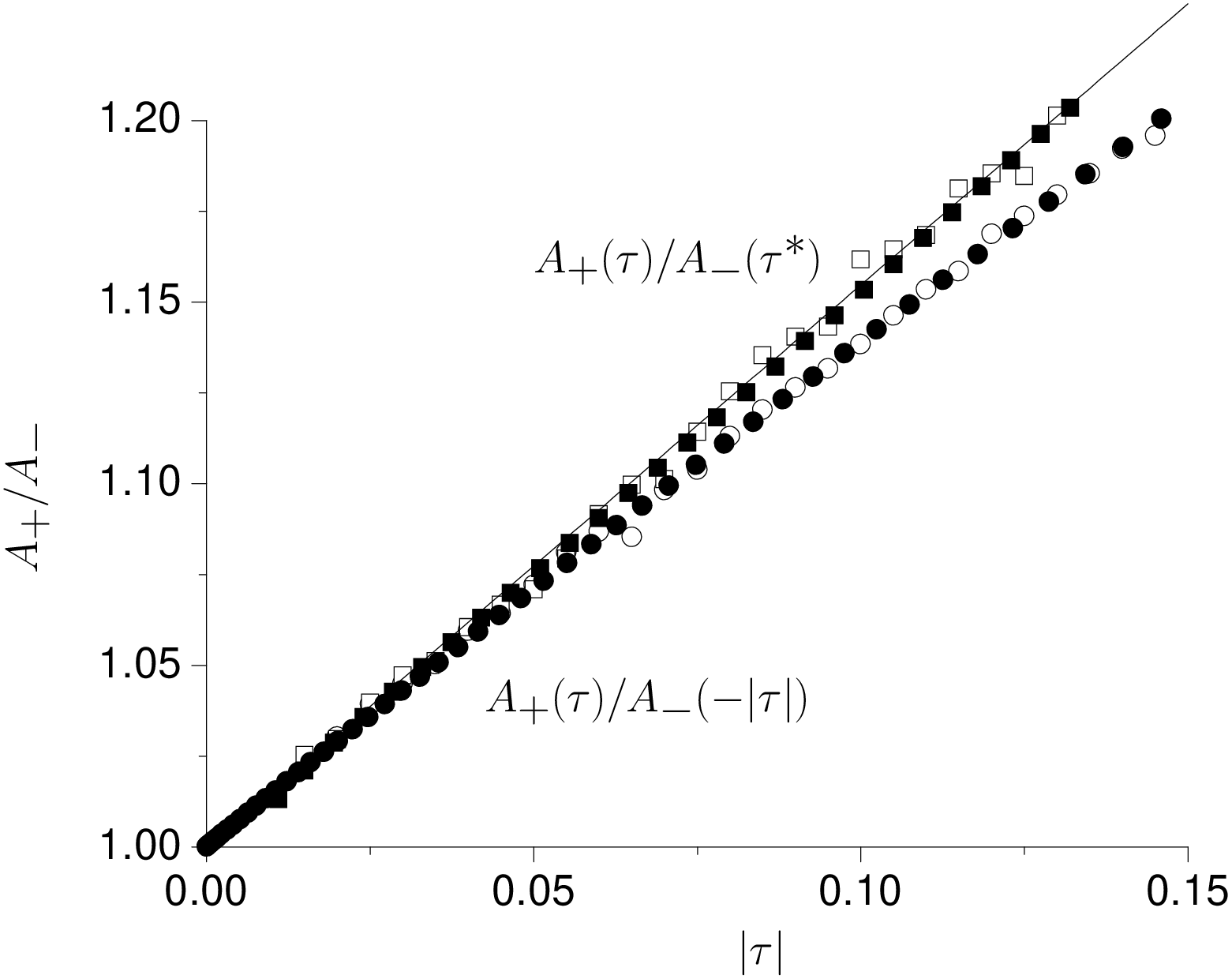,width=0.60\textwidth}
    \vskip-11.4cm
    \hskip8.7cm
  \epsfig{file=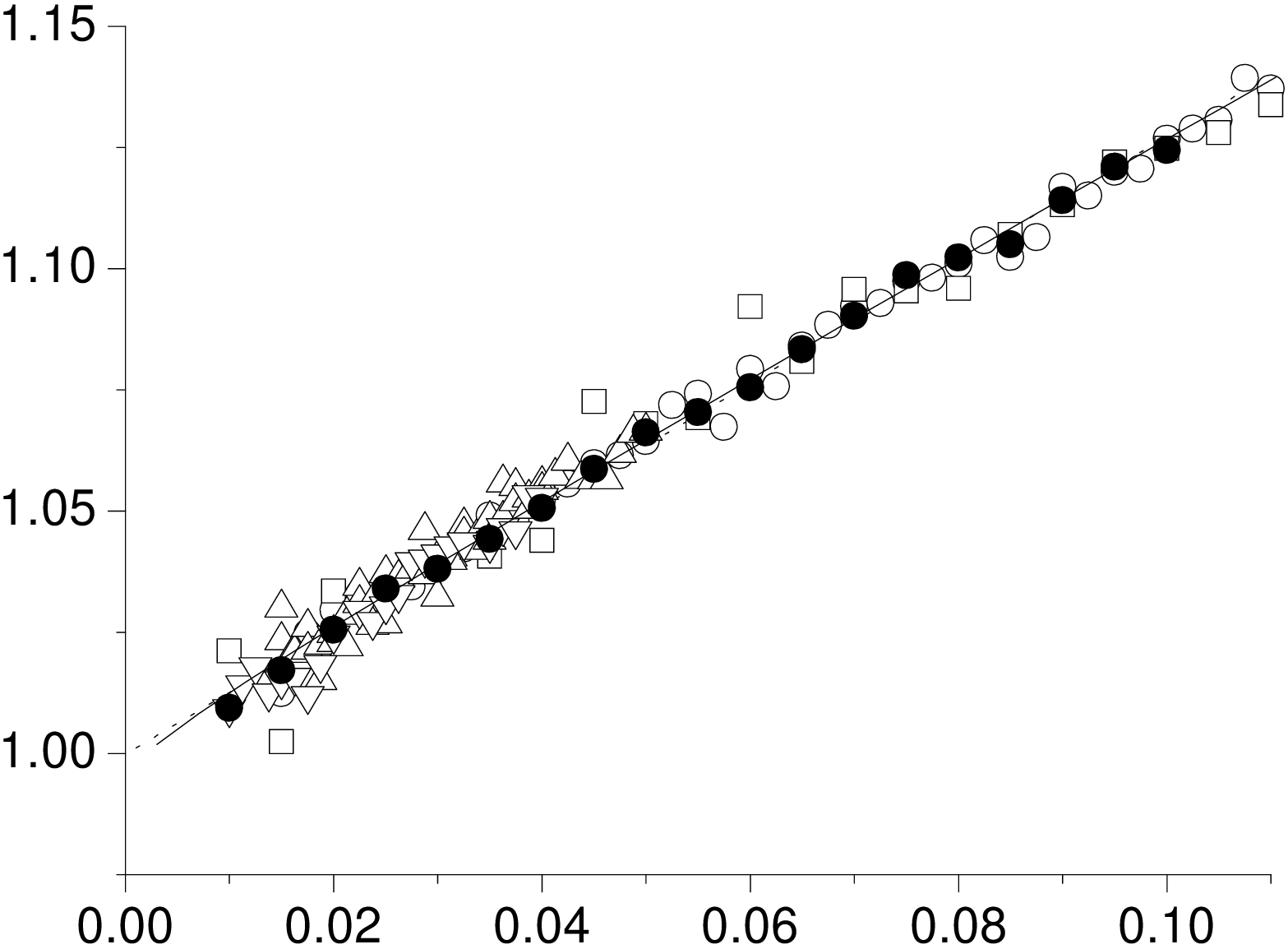,width=0.34\textwidth}
  \end{minipage}
  \vspace{0.8cm}
  \caption{\small    The specific-heat effective-amplitude ratio
    computed from the energy ratio
$\frac{(E(\beta)-E_0)\tau^{\alpha -1}}
{(E_0-E(\beta^*))(\tau^*)^{\alpha -1}}$
    as    a  function
    of the reduced temperature $\tau$ on the dual line.
    The SE data are represented by boxes,  our MC data
      ($L=100$)     by circles. The solid line is
    given by \Eq{d-e-linear}.
    The  ratio of the effective amplitudes $A_+(\tau)/A_-(-|\tau|)$
 is shown   for different sizes   in the insert
        (for lattice linear sizes $L=100$ (open circles),
$L=200$ (squares), $L=300$    (up triangles), and
    $L=400$ (down triangles). The simulations were performed
         with $N_{th}=10^5$ thermalization steps and
    $N_{MC}=10^6$ MC steps.
    The results of a simulation for $L=100$
        with $N_{th}=10^6$ and
    $N_{MC}=10^7$, are represented by closed circles.
    The SE data  are represented by a solid line.
        The dotted line represents a linear fit of the
    closed circles.
    }\label{eff-a-4}
    \label{e4-dual}
\end{figure}

We now turn to the study of the specific-heat amplitudes.
Figure~\ref{eff-a-4} shows the effective-amplitude ratio
$A_+(\tau)/A_-(-|\tau|)$ computed at dual and symmetric temperatures
following the same prescription as in the case of the 3-state Potts
model studied in paper~\cite{I}.
The same ratio computed from SE data is also shown for comparison.
The linear fit of the most accurate MC data (lattice size $L=100$
computed with $10^7$ measurement steps) yields for this ratio the
value $0.9999(1)$, which is remarkably close to the exact value 1
derived from duality.

%
%

The ratio
\begin{equation}
\frac{A_+(\tau)}{A_-(\tau^*)}= \frac{(E(\beta)-E_0)\tau^{\alpha -1}}
{(E_0-E(\beta^*))(\tau^*)^{\alpha -1}},
    \label{rat-e-dual}
\end{equation}
where the constant $E_0$ is the value of the energy at the transition
temperature, $E_0=E(\beta_c)=-1-1/\sqrt{q}$, when  expanded
close to the transition point leads to
\begin{eqnarray}
    \frac{A_+(\tau)}{A_-(\tau^*)}&=&1+\frac{7} {3}\alpha_q \tau
    +O(\tau^{1+\alpha})
    \label{d-e-linear}
\end{eqnarray}
with $\alpha_q=
-E_0\beta_ce^{-\beta_c}=\frac{\ln(1+\sqrt{q})}{\sqrt{q}}\approx
0.5493$. \Eq{d-e-linear} predicts an  asymptotically linear
$\tau-$dependence of the effective-amplitude ratio. This linear
dependence is observed in Fig.~\ref{e4-dual} which also confirms
that the leading logarithmic corrections in the scaling function
${\cal X}_{\it corr}(-\ln(|\tau|))$ asymptotically cancel in the
ratio $A_+(\tau)/A_-(\tau^*)$. (Of course this is true also for the
same quantity computed at symmetric temperatures, since $\tau^*
\approx |\tau |$ for small values of $\tau$). Figure~\ref{e4-dual}
compares the effective-amplitude ratio obtained from MC simulation
and SE data with \Eq{d-e-linear}. The slope of MC data is 1.25, very
close to the predicted value 1.28.

Now, we make the natural conjecture (proven in the Appendix in the
absence of background corrections) that the cancelation of the
leading logarithmic corrections will also occur for the other
ratios. For the leading log-correction, $-\ln\tmod$, and for the
next correction in $\ln(-\ln\tmod)/(-\ln\tmod)$, this can be shown
analytically from the RG as first indicated by Cardy, Nauenberg and
Scalapino in Ref.~\cite{CardyNauenbergScalapino80} and by Salas and
Sokal in Ref.~\cite{SalasSokal97}. Our statement is stronger since
it extends also to the higher order log-terms, such as the next
correction in $1/(-\ln\tmod)$. We believe that
equations~(\ref{ap-eq100}) to (\ref{ap-eq101}) in the appendix are
exact, and since all the log-terms come from \Eq{ap-eq99}, they
should cancel in the appropriate ratios (i.e.  when the same powers
of the dilution field appear in the numerator and the denominator.
This is always the case when one considers  an effective
combination
tending to a universal ratio as $\tau\to 0$).

According to a RG analysis (see Appendix~\ref{ap1}), we may write the
energy in the critical region as
\begin{eqnarray}
    E_\pm(\pm\tmod)&=&
    E_0\pm \frac{A_\pm}{\alpha(1-\alpha)\beta_c}
   \frac{\tmod^{1/3}}{(-\ln\tmod)}
    \frac{1+a_\pm |\tau|^{2/3}
    }{{\cal E}(-\ln\tmod){\cal F}(-\ln\tmod)}+
    D_{1,\pm} \tmod
    .
    \label{en-4s}
\end{eqnarray}
In a fixed range of values of the reduced temperature, the
``correction function'' ${\cal F}(-\ln|\tau|)$ is now fixed and the
only remaining freedom is to include background terms   (coeff. $D$)
and possibly additive corrections to scaling coming from irrelevant
scaling fields   (coeff. $a$). Therefore, once the function ${\cal
F}(-\ln\tmod)$ is fixed after our study of the magnetization, a
reasonable fit of the energy data needs only three
parameters\footnote{Like in the case of the magnetization, a fourth
parameter in $b_\pm\tmod$ appears to be unnecessary.}, $A_\pm$,
$a_\pm$, and $D_\pm$. The next step is the fit of the mean $\bar
A(\tau)$ of
 the effective amplitudes,
\begin{eqnarray}
    \bar A(\tau)&=&\frac 12\alpha(1-\alpha)\beta_c
        (E_+(\tau)-E_-(-|\tau|))\times {\cal G}(-\ln\tmod)/\tmod^{1-\alpha}
    \nonumber\\
    &=& \frac {\beta_c} {9} (E_+(\tau)-E_-(-|\tau|))
   \times {\cal G}(-\ln\tmod)/\tmod^{1/3}.
    \label{e'4}
\end{eqnarray}
We thus fit the MC data to the expression
\begin{equation}
    \bar A(\tau)=A (1+a \tmod^{2/3})
    +
    D_1 \tmod
    \times {\cal G}(-\ln\tmod)/\tmod^{1/3}.
    \label{e4-fit}
\end{equation}
The log corrections, which indeed cancel in the singular part,
unfortunately reappear in  the background term, albeit
suppressed by the power $ \tmod^{2/3}$.

\begin{table}[ht]
\caption{\small Fits of the energy difference (MC data computed at
{\em dual} temperatures) to the expression $\bar A(\tau)\simeq {\rm
Ampl.}\times(1+{\rm corr.\ terms})
     +{\rm backgrd.\ terms}\times \tmod^{-1/3}{\cal G}(-\ln\tmod)$.
    }
\center\scriptsize
\begin{tabular}{llllll}
\hline\noalign{\vskip-1pt}\hline\noalign{\vskip2pt} Fit \# &
Amplitude & \multispan{3}{\hfil correction terms\hfil} &
\multispan{1}{\hfil background term\hfil}
\\
\multispan{1}{\vrule height0pt width0mm} &\multispan{1}{\vrule
height0.5pt width17mm} & \multispan{3}{\vrule height0.5pt width54mm}
& \multispan{1}{\vrule height0.5pt width22mm}
\\
\noalign{\vskip2pt}
        & Ampl.  & $\propto\frac 1{-\ln|\tau|} $
        & $\propto\frac {\ln(-\ln\tmod)}{(-\ln|\tau|)^2} $
        & $\quad\propto|\tau|^{2/3}$& $\quad\propto|\tau|$\\
\noalign{\vskip2pt} \hline
MC \# $2^*$ & $1.338(3)$ & ${\bf -0.757}$ & ${\bf -0.522}$ & $-4.98(8)$ & $\phantom- 0.920(13)$\\
MC \# $3^*$ & $1.316(10)$ & ${\bf -0.88}$ & $\phantom- -$ & $-4.88(38)$ & $\phantom- 0.899(62)$ \\
\hline\noalign{\vskip-1pt}\hline\noalign{\vskip2pt}
\end{tabular}
\label{table_parameters_E}
\end{table}

\begin{figure}[ht]
  \centering
  \begin{minipage}{\textwidth}
  \epsfig{file=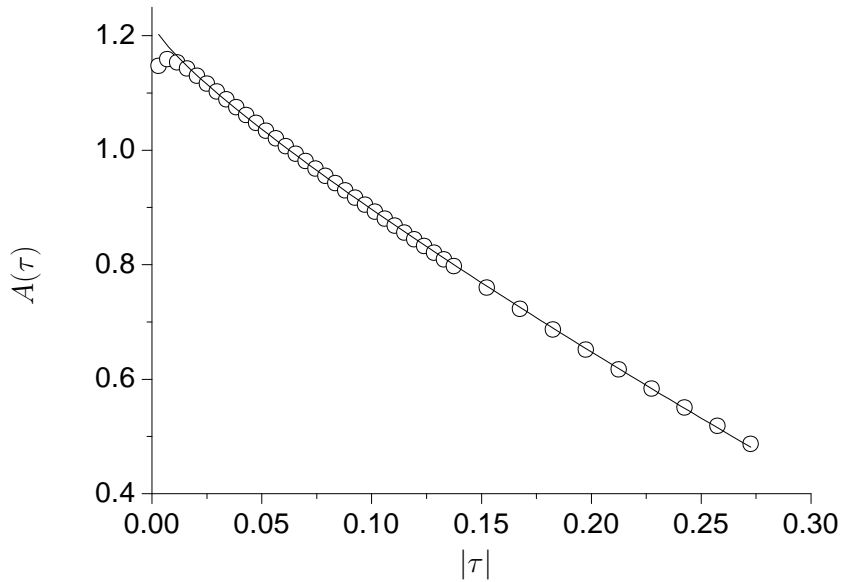,width=0.60\textwidth}
  \end{minipage}
  \vspace{-3cm}
\caption{\small The effective amplitude $\bar A(\tau)$ computed
    from our MC data using \Eq{e'4} (open circles) and from the fit
    to \Eq{e4-fit} (solid line).
}
    \label{e-eff-4}
\end{figure}
\begin{figure}[ht]
  \centering
  \begin{minipage}{\textwidth}
  \epsfig{file=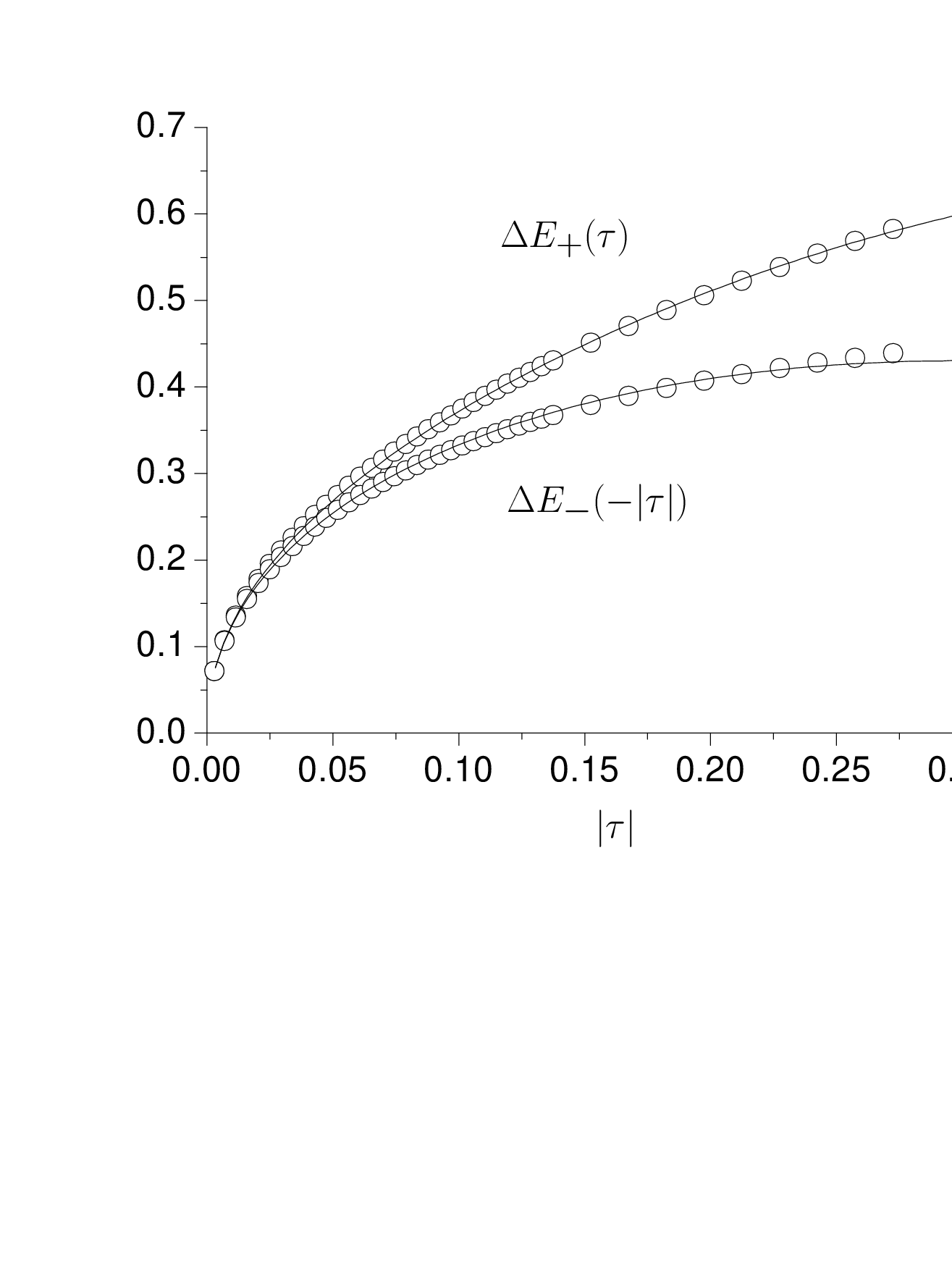,width=0.60\textwidth}
  \end{minipage}
  \vspace{-3cm}
\caption{\small
     The energy differences $\Delta E_+$ and
    $\Delta E_-$, calculated from our MC data for lattice size $L=100$
    (circles) and the fitted expressions (solid lines). }
    \label{e4-jan-symm}
\end{figure}

We note that $\bar A(\tau)$ is constructed in Eq.~(\ref{e'4})  using
the values of energy density computed at {\em symmetric}
temperatures. The same quantity constructed from the energy
densities at {\em dual} temperatures (with $E_-(\tau^*)$ instead of
$E_-(-\tmod)$ in Eq.~(\ref{e'4})) can also be studied and provides
better results. In table~\ref{table_parameters_E}, we show our
results (the best fit is obtained with the choices of $C_i$'s
coefficients labeled $2^*$ in tables~\ref{table_parameters_M} and
\ref{table_parameters_M-SE}). Again, the agreement between the two
fits is quite good, but this time it is more trivial since  the same
data set  is fitted.
 Taking the average of the parameters from the two fits, we conclude

\be A=1.327(12),\ee
and $a \simeq-4.93(23)$, and $D_1\simeq 0.910(38)$.
 Pad\'e approximants of SE data for the specific heat provide
$A=1.35(1)$.

 The expansion
including corrections to scaling and background terms for the
specific heat follows from the expressions of the energy density.
There is some disagreement between our amplitude
$\frac{A}{\alpha(1-\alpha)\beta_c}\simeq 5.922(40)-6.021(13)$ and the
result reported by Caselle et al.~\cite{CaselleTateoVinci99},
$6A\simeq 7.80(36)$\footnote{Notice that Caselle et al. use a
different definition of the energy.}.  We have to notice that the
amplitudes are very sensitive to the expression used for the fits.
Our choice of effective amplitude in \Eq{e'4} is supported by the
quite regular behavior  shown in figure~\ref{e-eff-4}, and also by
the natural choice of the fitting expression~(\ref{e4-fit}). The comparison
between the MC data and the resulting fit is shown in
figure~\ref{e4-jan-symm}. We
agree on this point with Enting and Guttmann~\cite{EntingGuttmann03}
who emphasized that their estimates depend {\em critically} on the
form assumed for the logarithmic sub-dominant terms, and on the
further assumption that the other sub-dominant terms,  including
powers of logarithms, powers of logarithms of logarithms etc., can
all be neglected.

%
%

\subsection{Susceptibilities amplitudes}\label{subsecChiq4}
\subsubsection{High temperature susceptibility amplitude}
We proceed
along the same lines as for the other physical quantities and fit
the high-temperature susceptibility to the expression
\begin{equation}
    \chi_+(\tau) =
    \Gamma_+\tau^{-7/6}{\cal G}^{3/4}(-\ln\tau)
    (1+a_{+}\tau^{2/3}+b_+\tau)
    +D_+.
    \label{chi-4-sing}
\end{equation}
%
\begin{figure}[ht]
  \centering
  \begin{minipage}{\textwidth}
  \hskip-2cm\epsfig{file=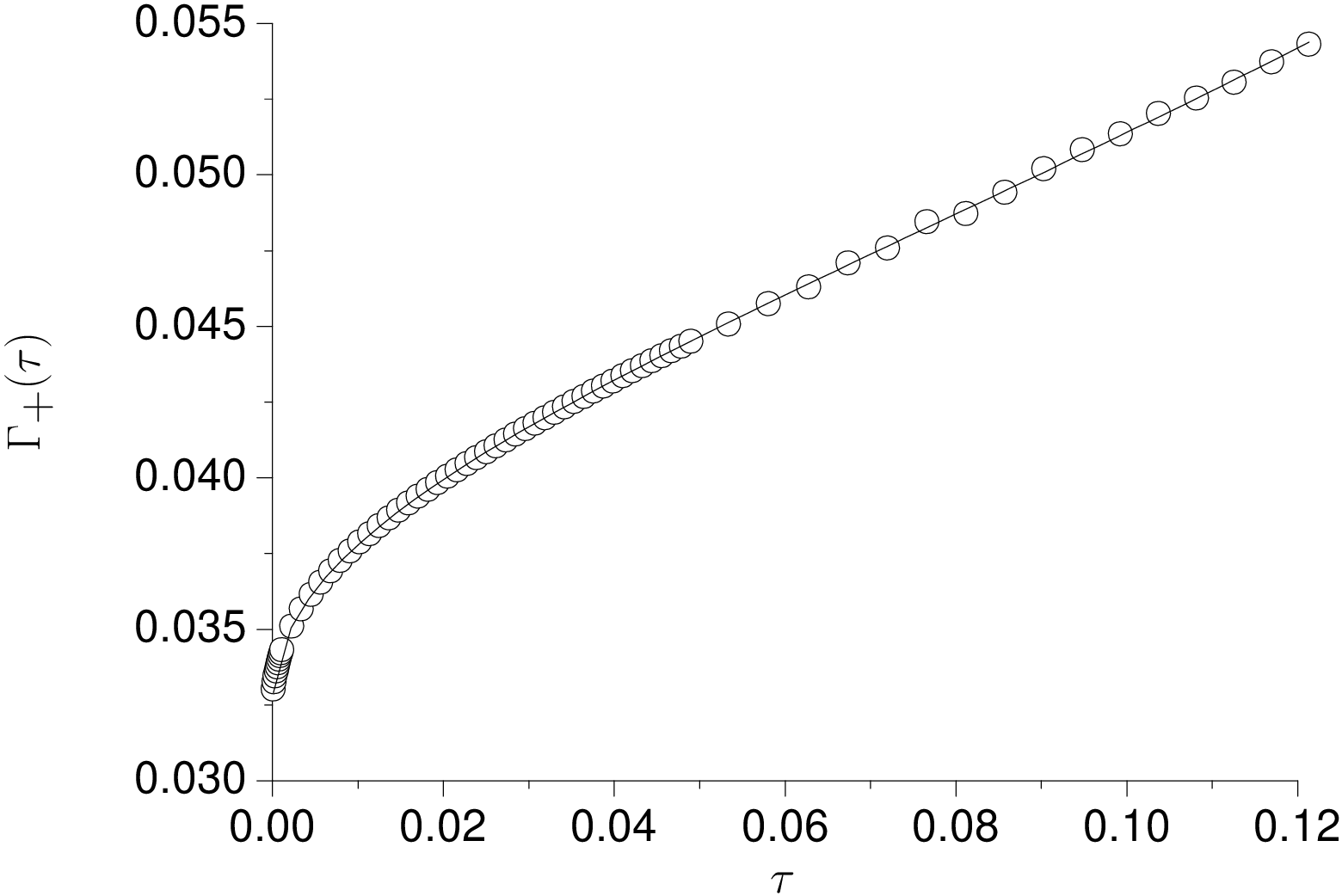,width=0.60\textwidth}
  \end{minipage}
  \vspace{-3cm}
\caption{\small The effective amplitude of the high-temperature
    susceptibility $\chi_+(\tau)$. We have shown SE data for $\tau\le 0.05$,
our MC data for $\tau> 0.05$ and a fit to \Eq{chi-4-sing} (dotted
    line).}
    \label{FigGamma+q4}
\end{figure}
\begin{table}[h]
\caption{\small Fits of the high-temperature susceptibility  to the
expression $\chi_+(\tau)\simeq {\rm Ampl.}\times\tau^{-7/6}
    \times {\cal G}^{3/4}(-\ln\tau)\times(1+{\rm corr.\ terms})
    +{\rm backgr.\ terms}$. A star in the first column indicates
  that the coefficients $C_i$'s are those
deduced from the MC fits of
table~\ref{table_parameters_M}.}
\center\scriptsize
\begin{tabular}{llllll}
\hline\noalign{\vskip-1pt}\hline\noalign{\vskip2pt} fit \# &
amplitude & \multispan{3}{\hfil correction terms \hfil} &
\multispan{1}{\hfil backgr.\hfil}
\\
\multispan{1}{\vrule height0pt width0mm} & \multispan{1}{\vrule
height0.5pt width17mm} & \multispan{3}{\vrule height0.5pt width50mm}
&\multispan{1}{\vrule height0.5pt width20mm}
\\
\noalign{\vskip2pt}
        & $\Gamma_+$  & $\propto\frac 1{-\ln\tau} $
        & $\propto\frac {\ln(-\ln\tau)}{(-\ln\tau)^2} $
    & $\quad\propto\tau^{2/3}$
    & $\quad\propto\tau^0$
       \\
\noalign{\vskip2pt} \hline
MC \# $2^*$ & $0.03144(15)$ & $\bf -0.757$ & $\bf -0.522$ & $\phantom- 0.561(60)$
    & $-0.053(17)$ \\
MC \# $3^*$ & $0.03178(30)$ & $\bf -0.88$ & $\phantom- -$ & $\phantom- 0.53(23)$
    & $\phantom- 0.052(120)$ \\
CTV \# $3^*$ &  $0.03051(29)$ &  ${\bf -0.88}$ & $\phantom - -$ & $\phantom- 1.48(34)$ & $-0.45(24)$\\
SE \# $2^*$ & $0.03041(1)$ & ${\bf -0.757}$ & $\bf -0.522$ & $\phantom- 1.30(1)$
    & $-0.362(9)$ \\
SE \# $3^*$ & $0.03039(1)$ & $\bf -0.88$ & $\phantom- -$ & $\phantom- 1.67(1)$
    & $-0.59(2)$ \\
 \hline\noalign{\vskip-1pt}\hline\noalign{\vskip2pt}
\end{tabular}
\label{table_parameters_Chi+}
\end{table}

We can easily obtain the amplitude $\Gamma_+$ observing that a
single constant as a background term $D_+$ is sufficient for the fit
(this will also be the case at low temperature). The effective
amplitude
$$\Gamma_{eff}(\tau)=\chi_+(\tau)\tau^{7/6}{\cal G}^{-3/4}(-\ln\tau)$$
is represented in Fig.~\ref{FigGamma+q4} and
table~\ref{table_parameters_Chi+} collects the coefficients
determined by the fits. We finally obtain the high-temperature
susceptibility amplitude
  \be\Gamma_+=0.0310(7).\ee
The value which
follows from
differential approximants to SE data, although less
accurate, is consistent with it, $\Gamma_+=0.033(2)$.

\subsubsection{Low temperature susceptibilities amplitudes}
 The behavior of the
longitudinal susceptibility in the low-temperature phase
 is less easy to analyze~\cite{SBB_EPL}.
We use the expression
\begin{equation}
    \chi_L(-\tmod) =
    \Gamma_L\tmod^{-7/6}{\cal G}(-\ln\tmod)^{3/4}
    (1+a_{L}\tmod^{2/3}+b_L\tmod)
    +D_L.
    \label{chiL-4-sing}
\end{equation}
and the various coefficients are collected in
table~\ref{table_parameters_ChiLandT}.
For the transverse susceptibility, the same procedure leads to the
amplitudes also listed in the table~\ref{table_parameters_ChiLandT}.

One may note that the values of the transverse susceptibility amplitude
are more stable than those of the
longitudinal amplitude, while the estimates of the
corrections to scaling are less scattered in the latter case.
Our final estimates are
  \be \Gamma_L=0.00478(24)\ee   and
\be\Gamma_T=0.00074(2).\ee DA analysis of SE data gives
approximately $\Gamma_L=0.005(1)$.

\begin{table}[ht]
\caption{\small Fits of the low-temperature longitudinal
susceptibility  to the expression $\chi_L(-\tmod)\simeq {\rm
Ampl.}\times\tmod^{-7/6}
    \times {\cal G}^{3/4}(-\ln\tmod)\times(1+{\rm corr.\ terms})
    +{\rm backgr.\ terms}$ and of the low-temperature transverse
susceptibility to the expression $\chi_T(-\tmod)\simeq {\rm
Ampl.}\times\tmod^{-7/6}
    \times {\cal G}^{3/4}(-\ln\tmod)\times(1+{\rm corr.\ terms})
    +{\rm backgr.\ terms}$. A star
in the first column indicates our favorite fit with the
coefficients deduced from the MC fits of
table~\ref{table_parameters_M}.}
\center\scriptsize
\begin{tabular}{llllll}
\hline\noalign{\vskip-1pt}\hline\noalign{\vskip2pt} fit \# &
amplitude & \multispan{3}{\hfil correction terms \hfil} &
\multispan{1}{\hfil backgr.\hfil}
\\
\multispan{1}{\vrule height0pt width0mm} & \multispan{1}{\vrule
height0.5pt width17mm} & \multispan{3}{\vrule height0.5pt width50mm}
&\multispan{1}{\vrule height0.5pt width20mm}
\\
\noalign{\vskip2pt}
        & $\Gamma_L$  & $\propto\frac 1{-\ln\tau} $
        & $\propto\frac {\ln(-\ln\tau)}{(-\ln\tau)^2} $
    & $\quad\propto\tau^{2/3}$
    & $\quad\propto\tau^0$
       \\
\noalign{\vskip2pt} \hline
MC \# $2^*$ & $0.00454(2)$ & $\bf -0.757$ & $\bf -0.522$ & $-2.83(3)$
    & $\phantom- 0.050(2)$ \\
MC \# $3^*$ & $0.00484(3)$ & $\bf -0.88$ & $\phantom- -$ & $-3.73(14)$
    & $\phantom- 0.13(1)$ \\
CTV \# $3^*$ &  $0.00494(3)$ &  $\bf -0.88$ & $\phantom - -$ & $ -4.35(15)$ & $\phantom- 0.210(19)$\\
SE \# $2^*$ & $0.00483(1)$ & $\bf -0.757$ & $\bf -0.522$ & $-3.77(3)$
    & $\phantom- 0.116(3)$ \\
SE \# $3^*$ & $0.00493(1)$ & $\bf -0.88$ & $\phantom- -$ & $-4.18(5)$
    & $\phantom- 0.178(6)$ \\
\noalign{\vskip2pt} \hline
\hline\noalign{\vskip-1pt}\hline\noalign{\vskip2pt}
        & $\Gamma_T$  & $\propto\frac 1{-\ln\tau} $
        & $\propto\frac {\ln(-\ln\tau)}{(-\ln\tau)^2} $
    & $\quad\propto\tau^{2/3}$
    & $\quad\propto\tau^0$
\\
\noalign{\vskip2pt}\hline\noalign{\vskip2pt}
MC \# $2^*$ & $0.00076(1)$ & $\bf -0.757$ & $\bf -0.522$ & $-0.805(34)$
    & $-0.0028(2)$ \\
MC \# $3^*$ & $0.00073(1)$ & $\bf -0.88$ & $\phantom- -$ & $-0.25(13)$
    & $-0.0050(14)$ \\

SE \# $2^*$ & $0.00073(1)$ & $\bf -0.757$ & $\bf -0.522$ & $-0.577(14)$
    & $-0.00373(15)$ \\
SE \# $3^*$ & $0.00073(1)$ & $\bf -0.88$ & $\phantom- -$ & $-0.369(15)$
    & $-0.00457(16)$ \\
\noalign{\vskip2pt} \hline
\hline\noalign{\vskip-1pt}\hline\noalign{\vskip2pt}
\end{tabular}
\label{table_parameters_ChiLandT}
\end{table}


\subsection{Universal amplitude ratio $R_c^-$.}

\begin{figure}[h]
  \centering
  \begin{minipage}{\textwidth}
  \hspace{-1cm}
  \epsfig{file=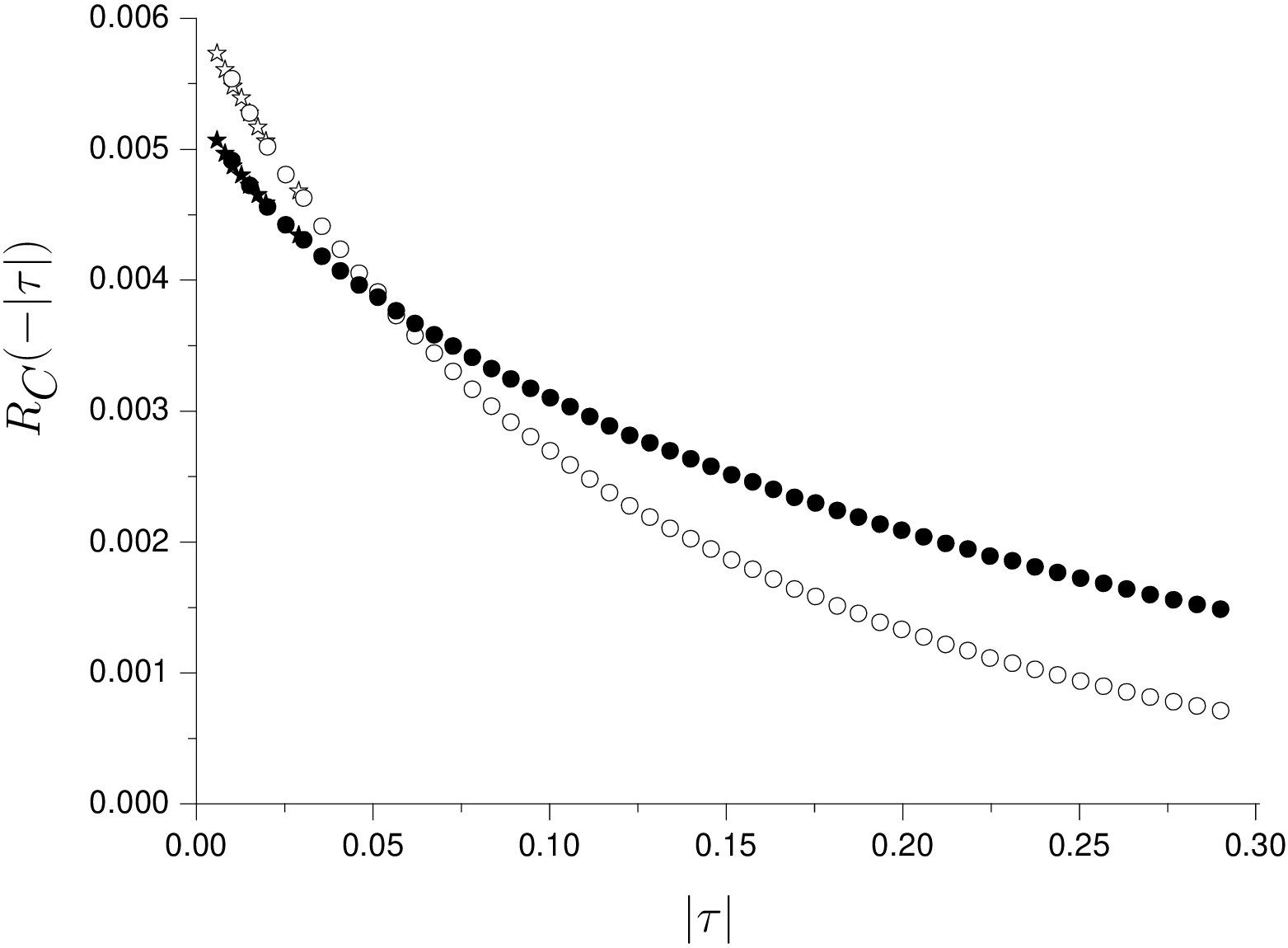,width=0.60\textwidth}
  \end{minipage}
  \vspace{-3cm}
  \caption{\small The functions ${R}^-_C(-\tmod)$ (open symbols)
    and ${R}^-_{C^*}(-\tau)$
  (closed symbols) which approaches the universal ratio
  $R_C^-$ as $\tau \to 0^-$. Our MC data are represented by circles  and the CTV data by stars.}
    \label{fig-rat-m}
\end{figure}

We use the available MC data for $C$, $M$, and $\chi$ to estimate the
universal amplitude ratio $R_C^-$ in the low-temperature phase.
To this purpose,
we form the function (compare with \Eq{rat-app})
\begin{equation}
R_C^-(-\tmod)=\alpha\tau^2\frac{C(-\tmod)\chi_L(-\tmod)}{M^2(-\tmod)}
\label{func-rat}
\end{equation}
which is an estimator of the universal amplitude ratio in the limit
$\tmod\rightarrow 0$. As discussed in the Appendix, we expect that
all sets of logarithmic corrections cancel in this ratio.
Figure~\ref{fig-rat-m} shows with open symbols the combination from
\Eq{func-rat} for two sets of MC data, those of Caselle et al. (CTV)
and our simulations. We may fit these data with
correction-to-scaling terms starting from $\tmod^{2/3}$ or, assuming
in  plain analogy with the energy ratio that such corrections
cancel, with terms starting with $\tmod$. In table~\ref{t-rat-m} we
include these fits for our MC data set varying the temperature
window and the number of correction terms.

\begin{table}[ht]
\caption{Estimates of the critical amplitude ratio
$R_C^-$ from our MC data using different fits and varying the temperature
window.}
\center\scriptsize
\begin{tabular}{llllll}
\hline\noalign{\vskip-1pt}\hline\noalign{\vskip2pt}
$\tau-$window & amplitude & \multispan{4}{\hfil correction terms \hfil} \\
\multispan{1}{\vrule height0pt width0mm} & \multispan{1}{\vrule
height0.5pt width17mm} & \multispan{4}{\vrule height0.5pt width80mm}\\
\noalign{\vskip2pt}
    & $R_C^-$  & $\quad\propto\tmod^{2/3}$ & $\quad\propto\tmod$
    & $\quad\propto\tmod^{4/3}$ & $\quad\propto|\tau|^{5/3}$
       \\
\noalign{\vskip2pt}\hline\noalign{\vskip2pt}
$0.01-0.29$
 & 0.00651(3) &$-3.04(12)$ & $-1.24(39)$ & $\phantom - 4.18(32)$ & $\phantom --$ \\
 & 0.00685(3) & $-4.60(3)$ & $\phantom -3.84(5)$ & $\phantom --$ & $\phantom --$ \\
\cline{2-6}\\ \noalign{\vskip-7pt}
 & 0.00619(1)& $\phantom --$ & $-16.13(11)$ & $\phantom -29.25(37)$ & $-14.34(31)$ \\
 & 0.00591(2) & $\phantom --$ & $-11.01(10)$ & $\phantom -12.13(16)$ & $\phantom --$\\
\noalign{\vskip2pt}\hline\noalign{\vskip2pt}
$ 0.01-0.10$
 & 0.00628(7) & $-1.54(58)$ & $-6.78\pm 2.36$ & $\phantom -9.41\pm 2.53$ & $\phantom --$ \\
 & 0.00654(3) & $-3.66(9)$ & $-1.99(18)$ & $\phantom --$ & $\phantom --$ \\
 & 0.00627(3) & $-2.70(2)$ & $\phantom --$ & $\phantom --$ & $\phantom --$ \\
\cline{2-6}\\ \noalign{\vskip-7pt}
 & 0.00618(3)& $\phantom --$ & $-15.90\pm 1.26$ & $\phantom -28.64\pm 5.51$ & $-14.09\pm 6.18$ \\
 & 0.00611(1) & $\phantom --$ & $-13.05(17)$ & $\phantom -16.10(35)$ & $\phantom --$\\
\noalign{\vskip2pt}\hline\noalign{\vskip2pt}
$0.01-0.046$
 & 0.00642(6) & $-3.09(33)$ & $\phantom -0.62(80)$ & $\phantom --$ & $\phantom --$ \\
 & 0.00637(1) & $-2.83(2)$ & $\phantom --$ & $\phantom --$ & $\phantom --$ \\
\cline{2-6}\\ \noalign{\vskip-7pt}
 & 0.00616(4)& $\phantom --$ & $-14.20(94)$ & $\phantom -18.99\pm 2.44$ & $\phantom --$ \\
 & 0.00588(4) & $\phantom --$ & $-6.92(18)$ & $\phantom --$ & $\phantom --$\\
\hline\noalign{\vskip-1pt}\hline\noalign{\vskip2pt}
\end{tabular}
\label{t-rat-m}
\end{table}
A similar analysis was also performed for the data from Caselle et al
(not shown here). The results of both analysis are
consistent.

More traditionally, we may evaluate the ratio $R_C^-$ using the
estimated values of the amplitudes reported in
sections~\ref{subsecMq4}, \ref{subsecEq4} and \ref{subsecChiq4} (see
\Eq{univ-rat}). The results will be presented later, but
anticipating the forthcoming analysis, we can quote as a  reliable
estimate $R_C^-\simeq 0.0052\pm 0.0002$   (approximants of SE data
lead to $0.0050(2)$). One can see from the table~\ref{t-rat-m} that
the ratio obtained from the effective function~\Eq{func-rat} is
systematically larger. The difference may be explained by the fact
that fitting the effective ratio function \Eq{func-rat} without any
logarithmic correction, we assume that the background corrections
for the longitudinal susceptibility $\chi_L$ and for the specific
heat $C$  are small in the critical temperature window. While this
is indeed the case, these background terms are not negligibly small
and their presence leads to systematic deviations of the estimates
of $R_C^-$ presented in the table~\ref{t-rat-m}.  The
ratio~(\ref{func-rat}) may be written as
$\alpha\tau^2(C_s+C_{bt})(\chi_s+\chi_{bt})/M^2$, where $C_s$ and
$\chi_s$ are the singular parts of the specific heat and of the
susceptibility and $C_{bt}$ and $\chi_{bt}$ are the corresponding
background nonsingular terms. \Eq{func-rat} may be rewritten as
$\alpha\tau^2 C_s\chi_s(1+C_{bt}/C_s)(1+\chi_{bt}/\chi_{s})/M^2$.
Thus, the background terms $C_{bt}$ and $\chi_{bt}$ contribute when
divided by the singular terms (or in other words multiplied by
factors $\tau^{2/3}{\cal G}$ and $\tau^{7/6}{\cal G}^{-3/4}$,
respectively). Clearly, in the critical region, the first factor has
the dominant contribution. This ``large'' term may be eliminated
completely if we form a quantity equivalent to \Eq{func-rat} from
the energy
 difference $E_-(\tmod)-E_0$ instead
of the specific heat,
\begin{equation}
R^{-}_{C^*}(-\tmod)=
\alpha(\alpha-1)\beta_c
\tau\frac{(E_-(-\tmod)-E_0)\chi_L(-\tmod)}{M^2(-\tmod)},
\label{func-rat-E}
\end{equation}
which is shown in the Figure~\ref{fig-rat-m} with closed symbols.
The extrapolation at $\tau\to 0^-$ is obviously different. The
results of the fit of MC data to \Eq{func-rat-E} are given  in
table~\ref{t-rat-m-e}. The outcome for the universal combination
$R_C^-$ is now fully consistent with the value $0.0052(2)$  and
supports our idea that the specific heat background term spoils the
behavior of the estimator~(\ref{func-rat}).

\begin{table}[ht]
\caption{Estimates of the critical amplitude ratio $R_{C^*}^-$
(Expr.~(\ref{func-rat-E})) from our MC data using different fits and
varying the temperature window.} \center\scriptsize
\begin{tabular}{llllll}
\hline\noalign{\vskip-1pt}\hline\noalign{\vskip2pt}
$\tau-$window & amplitude & \multispan{4}{\hfil correction terms \hfil} \\
\multispan{1}{\vrule height0pt width0mm} & \multispan{1}{\vrule
height0.5pt width17mm} & \multispan{4}{\vrule height0.5pt width80mm}\\
\noalign{\vskip2pt}
    & $R_C^-$  & $\quad\propto\tmod^{2/3}$ & $\quad\propto\tmod$
    & $\quad\propto\tmod^{4/3}$ & $\quad\propto|\tau|^{5/3}$
       \\
\noalign{\vskip2pt}\hline\noalign{\vskip2pt} $0.01-0.29$
 & 0.00551(9) &$-2.53(5)$ & $\phantom- 0.57(15)$ & $\phantom-  1.11(12)$ & $\phantom --$ \\
 & 0.00558(3) & $-2.97(3)$ & $\phantom- 1.95(2)$ & $\phantom --$ & $\phantom --$ \\
\cline{2-6}\\ \noalign{\vskip-7pt}
 & 0.00527(8)& $\phantom --$ & $-11.44(16)$ & $\phantom- 20.98(51)$ & $-11.25(42)$ \\
 & 0.00508(1) & $\phantom --$ & $-7.29(9)$ & $\phantom- 7.38(14)$ & $\phantom --$\\
\noalign{\vskip2pt}\hline\noalign{\vskip2pt} $ 0.01-0.10$
 & 0.00558(4) & $-3.28(36)$ & $\phantom- 3.66\pm 1.45$ & $\phantom -2.23\pm 1.55$ & $\phantom --$ \\
 & 0.00553(1) & $-2.77(5)$ & $\phantom- 1.58(9)$ & $\phantom --$ & $\phantom --$ \\
 & 0.00535(2) & $-1.98(2)$ & $\phantom --$ & $\phantom --$ & $\phantom --$ \\
\cline{2-6}\\ \noalign{\vskip-7pt}
 & 0.00538(2)& $\phantom --$ & $-16.11(99)$ & $\phantom- 40.39\pm 4.30$ & $31.97\pm 4.81$ \\
 & 0.00525(2) & $\phantom --$ & $-9.55(24)$ & $\phantom- 11.82(50)$ & $\phantom --$\\
\noalign{\vskip2pt}\hline\noalign{\vskip2pt} $0.01-0.046$
 & 0.00558(3) & $-3.09(20)$ & $\phantom- 2.34(48)$ & $\phantom --$ & $\phantom --$ \\
 & 0.00543(2) & $-2.11(3)$ & $\phantom --$ & $\phantom --$ & $\phantom --$ \\
\cline{2-6}\\ \noalign{\vskip-7pt}
 & 0.00535(3)& $\phantom --$ & $-12.40(74)$ & $\phantom -18.95\pm 1.99$ & $\phantom --$ \\
 & 0.00511(4) & $\phantom --$ & $-5.04(19)$ & $\phantom --$ & $\phantom --$\\
\hline\noalign{\vskip-1pt}\hline\noalign{\vskip2pt}
\end{tabular}
\label{t-rat-m-e}
\end{table}

\begin{table}[ht]
\caption{Estimates of the critical amplitude ratio $R_{C^*}^-$ from
Caselle et al MC data using different fits and varying the
temperature window.} \center\scriptsize
\begin{tabular}{lllll}
\hline\noalign{\vskip-1pt}\hline\noalign{\vskip2pt}
$\tau-$window & amplitude & \multispan{3}{\hfil correction terms \hfil} \\
\multispan{1}{\vrule height0pt width0mm} & \multispan{1}{\vrule
height0.5pt width17mm} & \multispan{3}{\vrule height0.5pt width60mm}\\
\noalign{\vskip2pt}
    & $R_C^-$  & $\quad\propto\tmod^{2/3}$ & $\quad\propto\tmod$
    & $\quad\propto\tmod^{4/3}$
       \\
\noalign{\vskip2pt}\hline\noalign{\vskip2pt} $0.0058-0.029$
 & 0.00548(2) & $-2.59(17)$ & $1.27(48)$ & $\phantom --$ \\
 & 0.00543(1) & $-2.14(2)$ & $\phantom --$ & $\phantom --$  \\
\cline{2-5}\\ \noalign{\vskip-7pt}
 & 0.00535(1) & $\phantom --$ &  $-13.77(66)$ & $\phantom -21.79\pm 1.98$ \\
 & 0.00521(2) & $\phantom --$ & $-5.93(23)$ & $\phantom --$ \\
\hline\noalign{\vskip-1pt}\hline\noalign{\vskip2pt}
\end{tabular}
\label{t-rat-m-cas}
\end{table}

Again, a similar analysis of CTV data (see table~\ref{t-rat-m-cas})
leads to fully consistent results.

\section{Discussion}

\label{sec-discussion}
 Our final goal is the determination of some universal combinations of
amplitudes.  This can be done either directly from the values of the
amplitudes listed in the various tables of this paper, or also by
extrapolating the {\em effective ratios} to $\tau = 0$. Let us start
with an estimate obtained by the second method, and let us concentrate
on the most controversial amplitude ratios,   those of
the susceptibilities.
As we have just shown in the section on $R_C^-$, this method may
lead to systematic deviations if the background terms are
not handled with care.
Later on we shall discuss the ratios of the amplitudes
 listed in the previous tables.

The  ratio $\Gamma_+/\Gamma_L$ can be estimated from the ratio of the SE
of the susceptibility $\chi_+$ at high-temperature and of the
longitudinal susceptibility $\chi_L$ in the low-temperature phase.
We have again two options to form this ratio, either from quantities computed
 at temperatures symmetric with respect to the critical temperature
$T_c\pm\tau$ or at  inverse temperatures $\beta$ and $\beta^*$
related by the duality relation~(\ref{d-t}).
\begin{figure}[h]
  \centering
  \begin{minipage}{\textwidth}
  \epsfig{file=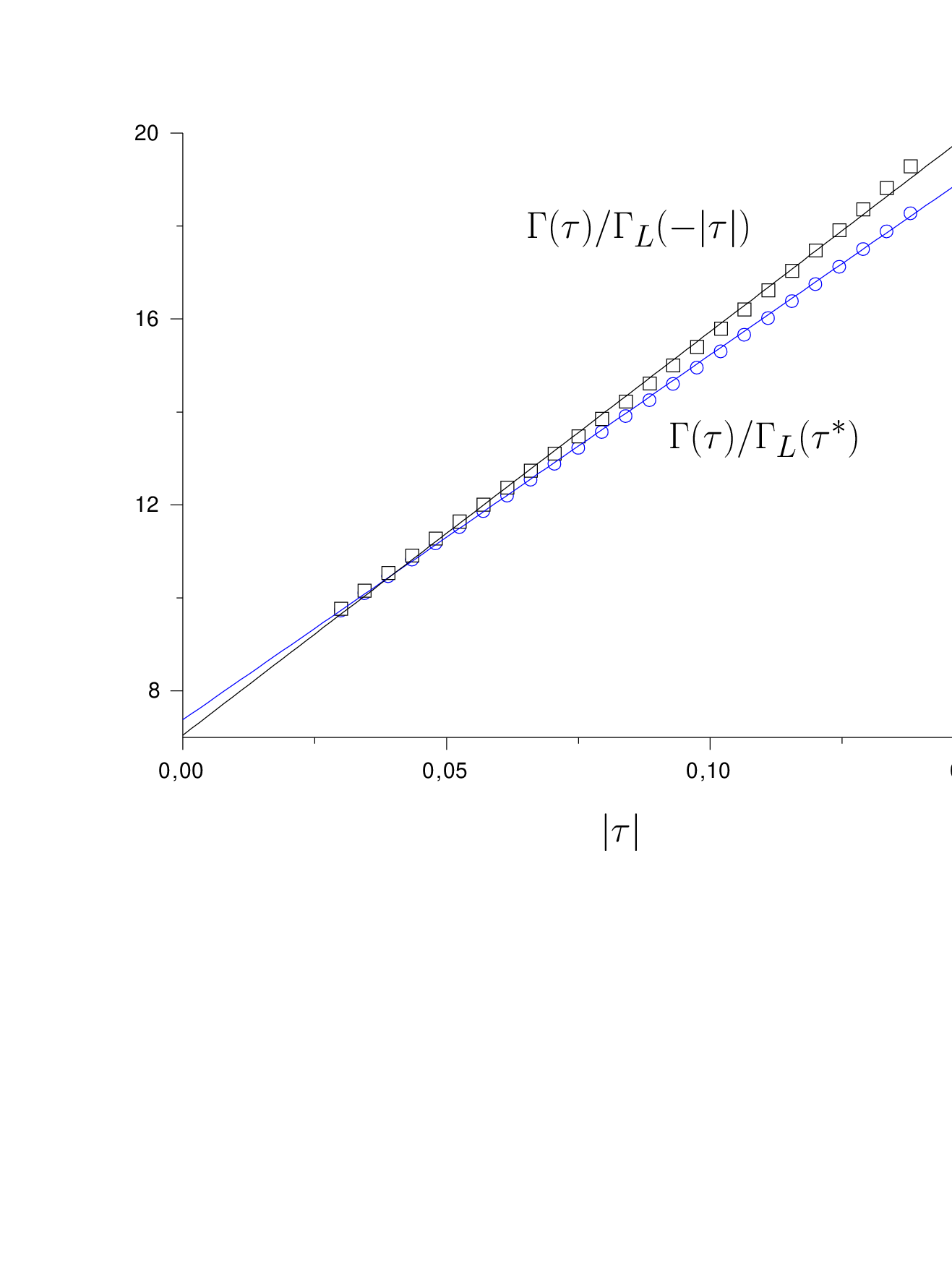,width=0.55\textwidth}
  \end{minipage}
  \vspace{-4.cm}
  \caption{\small The ratio of the effective amplitudes
    $\Gamma_+(\tau)/\Gamma_L(-\tmod)$
    obtained from
    SE data at symmetric temperatures (boxes) and dual temperatures
    (circles) together with the corresponding linear fits (solid lines).}
    \label{ggl-eff-4}
\end{figure}
Figure~\ref{ggl-eff-4} shows both the ratios
$\Gamma_+(\tau)/\Gamma_L(-\tmod)$ and
$\Gamma_+(\tau)/\Gamma_L(\tau^*)$, while the straight lines are drawn
as a guide for the eye. It would be naive to take the linear fit too
seriously, otherwise one should conclude that the background
(non-singular) contribution to the ratio is negligible. The value of the
universal amplitude ratio $\Gamma_+/\Gamma_L$ obtained from the SE
data is approximately $\Gamma_+/\Gamma_L\simeq 6.16(1)$ and $6.30(1)$,
when using respectively the fits MC $\# 2^*$ and MC $\# 3^*$. We have
also analyzed the effective-amplitude ratio from MC data obtained by
dividing the high-temperature reduced susceptibility by the
longitudinal reduced susceptibility computed at temperatures related
by the duality relation. Neglecting the constant background terms in
the susceptibilities eliminates all logarithms and makes the fit quite
simple, leading to a ratio in the range $\Gamma_+/\Gamma_L\simeq
6.30-6.60$.  On the other hand, if we keep in the fit the background
terms, the logs reappear and we are lead to $\Gamma_+/\Gamma_L\simeq
6.0-6.1$.

Thus we get values which are quite different from the analytical
prediction $\Gamma_+/\Gamma_L=4.013$ of Delfino and
Cardy~\cite{DelfinoCardy98}, as well as from the value $3.5(4)$
estimated by Enting and Guttmann from an analysis of the SE data for
the susceptibility in both phases, and from the value $3.14(70)$
estimated by Caselle et al.~\cite{CaselleTateoVinci99}.

\begin{figure}[ht]
  \centering
  \begin{minipage}{\textwidth}
  \epsfig{file=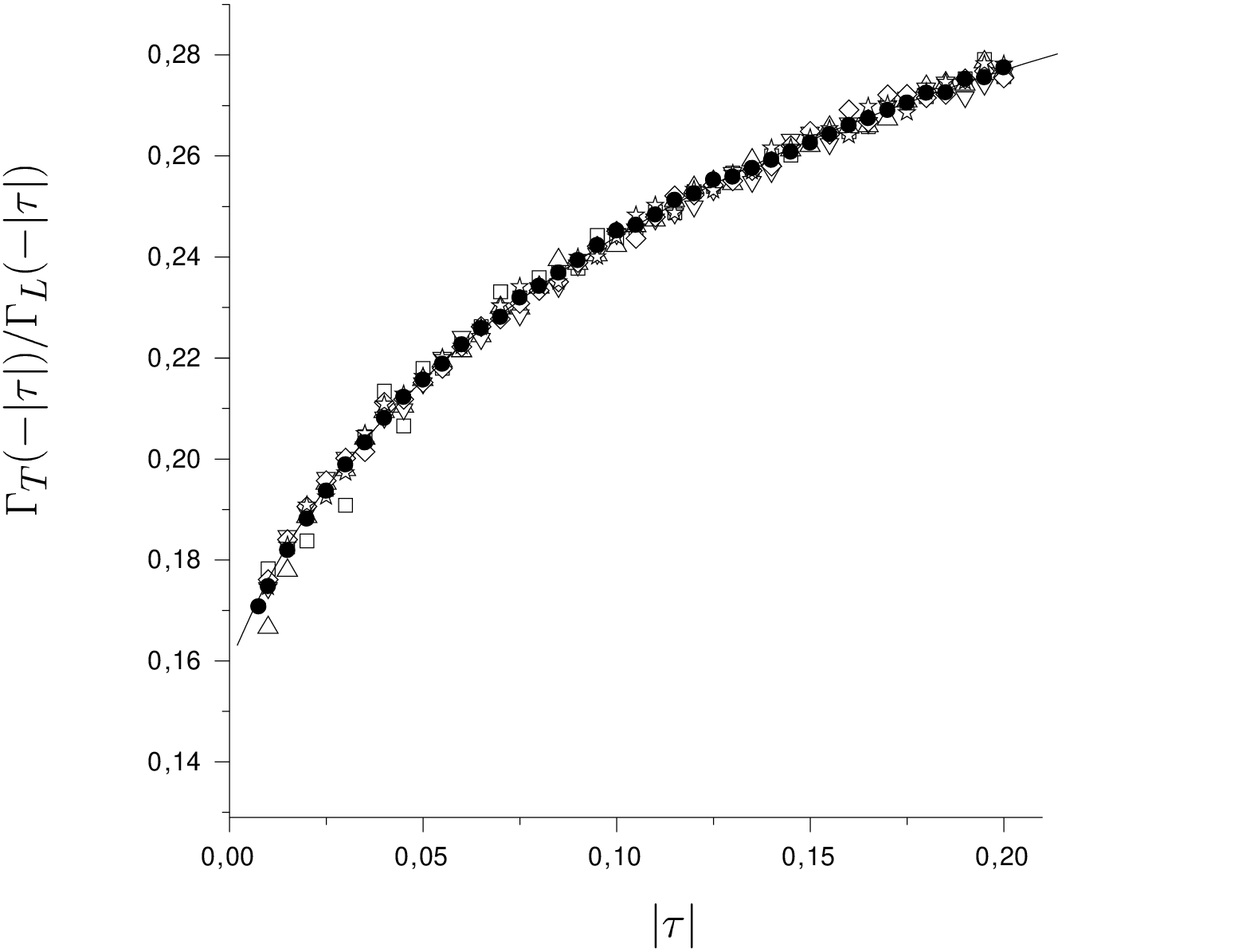,width=0.60\textwidth}
  \end{minipage}
  \vspace{-3.5cm}
  \caption{\small 4-state Potts model. The ratio of the effective amplitudes
    $\Gamma_T(-\tmod)/\Gamma_L(-\tmod)$ for the 4-state Potts model on
     square lattices of linear sizes $L=20$ (boxes), $L=40$ (up
    triangles), $L=60$ (down triangles), $L=80$ (diamonds), $L=100$
    (stars) computed with $N_{MC}=10^5$  MC steps, and
    $L=100$ (closed circles) computed with $N_{MC}=10^6$. The solid line
        represents the  SE data.}
    \label{gtgl-eff-4}
\end{figure}

Let us now estimate the effective-amplitude ratio
$\Gamma_T(-\tmod)/\Gamma_L(-\tmod)$.
This ratio, shown in figure~\ref{gtgl-eff-4}, has
been computed both by MC simulation, for various lattice sizes, and from
SE data. Due to the non-singular correction terms, its behavior is
far from being linear in $\tau$ as was the ratio
$\chi_+(\tmod)/\chi_L(-\tmod)$ (compare with
figure~\ref{ggl-eff-4}). A possible explanation is that there might be
some symmetry in the correction-to-scaling amplitudes occurring in the
asymptotic expansion of $\chi_+(\tmod)$ and $\chi_L(-\tmod)$, but not
of $\chi_T(-\tmod)$, which introduces here a stronger background term
still containing the logs. Following again the same procedure, we
arrive at estimates close to $0.146$ when neglecting all logarithmic
corrections, while we get $0.152-0.153$ when allowing for these corrections
 and from the analysis of SE data. Finally we obtain
$\Gamma_T/\Gamma_L=0.151(3)$ and $0.148(3)$ from the fits MC $\#2^*$
and $3^*$.  All these values differ from the analytical prediction
$\Gamma_T/\Gamma_L=0.129$ of Delfino et
al.~\cite{DelfinoBarkemaCardy00} and from the value $0.11(4)$ of
Ref.~\cite{EntingGuttmann03}.

Let us now extract what we believe are
 more reliable estimates for the universal
combinations of amplitudes by a direct evaluation of the ratios of the
numbers presented in this paper and collected again in
table~\ref{tab-q4-res}. The universal combinations are presented in
table~\ref{tab-3-resbisq4}, together with the corresponding results
available in the literature. Averaging our different results, we
quote the following final estimates:
\begin{eqnarray}
\Gamma_+/\Gamma_L&=&6.49\pm 0.44,\\
\Gamma_T/\Gamma_L&=&0.154\pm 0.012,\\
R_C^+&=&0.0338\pm 0.0009,\\
R_C^-&=&0.0052\pm 0.0002.
\end{eqnarray}
These results clearly confirm the
above-mentioned   limits of effective ratios.

\begin{table}[h]
\caption{\small Critical amplitudes and correction coefficients for
    the 4-state Potts model.  They are written in the following format
    ${\rm Obs.}(\pm\tmod)\simeq{\rm Ampl.}\times
    \tmod^{\blacktriangleleft}\times
    {\cal G}^{\bigstar}(-\ln\tmod)\times(1+{\rm corr.\ terms})
    +{\rm backg.\ terms}.$
    The results of the MC analysis
    of Ref.\cite{CaselleTateoVinci99}
    are compiled together with our results
    obtained  by combining the MC and series
    expansion (SE) data
    analysis.
} \center\scriptsize
\begin{tabular}{llllllr}
\hline\noalign{\vskip-1pt}\hline\noalign{\vskip2pt}
observable & amplitude &
\multispan{2}{\hfil correction terms, $\times {\cal G}^{\bigstar}(-\ln\tmod)$\hfil}
& \multispan{2}{\hfil background terms\hfil}
& source \\
\multispan{1}{\vrule height0pt width0mm} &
\multispan{1}{\vrule height0pt width0mm} &
\multispan{2}{\vrule height0.5pt width30mm} &
\multispan{2}{\vrule height0.5pt width32mm} &
\multispan{1}{\vrule height0pt width0mm}
\\
\noalign{\vskip2pt}
        &
    & $\quad\propto|\tau|^{2/3}$ & $\quad\propto|\tau|$
    & $\quad\propto|\tau|^0$ & $\quad\propto|\tau|$
       &        \\
\noalign{\vskip2pt} \hline
$E_\pm(\tau)$ & $1.338(3)/\alpha(1-\alpha)\beta_c$
    & $ -4.98(8)$ & $\phantom -\alpha_4$
    & $\phantom{--} E_0$ &  $\phantom -3.77(6)$
& this paper MC\# $2^*$ \\
& $1.316(9)/\alpha(1-\alpha)\beta_c$ 
    & $ -4.88(38)$ & $\phantom -\alpha_4$
    & $\phantom{--} E_0$ &  $\phantom -3.68(30)$
& this paper MC\# $3^*$\\
\hline
$\chi_+(\tau)$ & $\Gamma_+=0.0223(14)$ 
     & $\phantom{-}-$ & $\phantom{-}-$
    & $\phantom{-}0.05(14)$
    & $\phantom{-}-$
&\cite{CaselleTateoVinci99}\\
& $\Gamma_+=0.031(5)$ 
    & $\phantom{-}-$ & $\phantom{-}-$
    & $\phantom{-}-$
    & $\phantom{-}-$
&\cite{EntingGuttmann03}\\
& $\Gamma_+=0.03144(15)$ 
    & $\phantom{-}0.561(60)$ & $\phantom- -$
    & $ -0.053(17)$
    & $\phantom{-}-$
& this paper MC\# $2^*$\\
& $\Gamma_+=0.03178(30)$ 
    & $\phantom{-}0.53(23)$ & $\phantom- -$
    & $\phantom- 0.052(120)$
    & $\phantom{-}-$
& this paper MC\# $3^*$\\
& $\Gamma_+=0.03041(1)$ 
    & $\phantom{-}1.30(1)$ & $\phantom- -$
    & $ -0.362(9)$
    & $\phantom{-}-$
& this paper SE\# $2^*$\\
& $\Gamma_+=0.03039(1)$ 
    & $\phantom{-}1.67(1)$ & $\phantom- -$
    & $ -0.59(2)$
    & $\phantom{-}-$
& this paper SE\# $3^*$\\
\hline
$\chi_L(-|\tau|)$ & $\Gamma_L=0.00711(10)$ 
    & $\phantom{-}-$ & $\phantom{-}-$
    & $\phantom{-}0.02(1)$ & $\phantom{-}-$
& \cite{CaselleTateoVinci99}\\
& $\Gamma_L=0.0088(4)$ 
    & $\phantom{-}-$ & $\phantom{-}-$
    & $\phantom{-}-$
    & $\phantom{-}-$
&\cite{EntingGuttmann03}\\
& $\Gamma_L=0.00454(2)$ 
    & $-2.83(3)$ & $\phantom{-}-$
    & $\phantom- 0.050(2)$
    & $\phantom{-}-$
& this paper MC\# $2^*$\\
& $\Gamma_L=0.00484(3)$ 
    & $-3.73(14)$ & $\phantom{-}-$
    & $\phantom- 0.13(1)$
    & $\phantom{-}-$
& this paper MC\# $3^*$\\
& $\Gamma_L=0.00483(1)$ 
    & $-3.77(3)$ & $\phantom{-}-$
    & $\phantom- 0.116(3)$
    & $\phantom{-}-$
& this paper SE\# $2^*$\\
& $\Gamma_L=0.00493(1)$ 
    & $-4.18(5)$ & $\phantom{-}-$
    & $\phantom- 0.178(6)$
    & $\phantom{-}-$
& this paper SE\# $3^*$\\
\hline
$\chi_T(\tau)$ &   $\Gamma_T=0.0010(3)$   
    & $\phantom{-}-$ & $\phantom{-}-$
    & $\phantom{-}-$
    & $\phantom{-}-$
&\cite{EntingGuttmann03}\\
& $\Gamma_T=0.00076(1)$ 
    & $-0.805(34)$ & $\phantom{-}-$
    & $-0.0028(2)$
    & $\phantom--$
& this paper MC\# $2^*$\\
& $\Gamma_T=0.00073(1)$ 
    & $-0.25(13)$ & $\phantom{-}-$
    & $-0.0050(14)$
    & $\phantom--$
& this paper MC\# $3^*$\\
& $\Gamma_T=0.00073(1)$ 
    & $-0.577(14)$ & $\phantom{-}-$
    & $-0.00373(15)$
    & $\phantom--$
& this paper SE\# $2^*$\\
& $\Gamma_T=0.00073(1)$ 
    & $-0.369(15)$ & $\phantom{-}-$
    & $-0.00457(16)$
    & $\phantom--$
& this paper SE\# $3^*$\\
\hline
$M(-|\tau|)$ & $B=1.1621(11)$ 
    & $\phantom{-}-$ & $\phantom{-}-$
    & $\phantom{-}0.05(14)$ & $\phantom{-}-$
& \cite{CaselleTateoVinci99}\\
& $B=1.1570(1)$ 
    & $-0.191(2)$ & $\phantom{-}-$
    & $\phantom{-}-$ & $\phantom{-}-$
& this paper MC\# $2^*$\\
& $B=1.1559(12)$ 
    & $-0.210(10)$ & $\phantom{-}-$
    & $\phantom{-}-$ & $\phantom{-}-$
& this paper MC\# $3^*$\\
& $B=1.1575(1)$ 
    & $-0.194(1)$ & $\phantom{-}-$
    & $\phantom{-}-$ & $\phantom{-}-$
& this paper SE\# $2^*$\\
& $B=1.1575(1)$ 
    & $-0.225(1)$ & $\phantom{-}-$
    & $\phantom{-}-$ & $\phantom{-}-$
& this paper SE\# $3^*$\\
\hline\noalign{\vskip-1pt}\hline\noalign{\vskip2pt}
\end{tabular}
\label{tab-q4-res}
\end{table}

\begin{table}[h]
\caption{\small Universal combinations of the critical
amplitudes in the 4-state
    Potts model. }
\center\scriptsize
\begin{tabular}{lllllc}
\hline\noalign{\vskip-1pt}\hline\noalign{\vskip2pt} $A_+/A_-$ &
$\Gamma_+/\Gamma_L$ & $\Gamma_T/\Gamma_L$ & $R_C^+$ & $R_C^-$ &
source
\\ \noalign{\vskip2pt}\hline
   $1.$  &  $4.013$  & $0.129$  & $0.0204$  &  $0.00508$ & \cite{DelfinoCardy98,DelfinoBarkemaCardy00}\\
$\phantom --$   &  $4.02$       & $0.129$& $\phantom --$&$\phantom --$& \cite{DelfinoGrinza04}\\
$\phantom --$   &  $3.14(70)$& $\phantom --$ & $0.021(5)$&  $0.0068(9)$   & \cite{CaselleTateoVinci99}\\
$\phantom --$   &  $3.5(4)$       & $0.11(4)$& $\phantom --$&$\phantom --$& \cite{EntingGuttmann03}\\
$1.000(5)$    & $6.93(6)$ & $0.1674(30)$ & $0.03452(25)$ & $ 0.00499(3)$ & this paper MC\# $2^*$\\
$1.000(13)$   & $6.57(10)$& $0.1508(26)$ & $0.03439(63)$ & $ 0.00524(9)$ & this paper MC\# $3^*$\\
$\phantom --$ & $6.30(1)$ & $0.1511(24)$ & $0.03336(9)$ & $ 0.00530(3)$ & this paper SE\# $2^*$\\
$\phantom --$ & $6.16(1)$ & $0.1481(23)$ & $0.03279(24)$ & $ 0.00532(5)$ & this paper SE\# $3^*$\\
\hline\noalign{\vskip-1pt}\hline\noalign{\vskip2pt}
\end{tabular}
\label{tab-3-resbisq4}
\end{table}

The main outcome of this work are the surprisingly high values of the
ratios $\Gamma_+/\Gamma_L$, $\Gamma_T/\Gamma_L$ and $R_C^+$ (and the low
value for $R_C^-$), significantly deviating from the predictions of Delfino
and Cardy. We emphasize that our results are also supported by a direct
extrapolation of {\em effective-amplitude ratios} for which most
 corrections to scaling disappear.
We believe that our fitting procedure is reliable, and since the
disagreement with the theoretical calculations can hardly be resolved,
we suspect that it might be attributed to the
approximations made in Ref.~\cite{DelfinoCardy98} in order to predict
the susceptibility ratios. Even more puzzling is the fact that
Delfino and Cardy argue in favor of a higher robustness of their
results for $\Gamma_T/\Gamma_-$ than for $\Gamma_+/\Gamma_-$, while
the disagreement is indisputable in both cases. Indeed, in the
conclusion of their paper, and in a footnote, Delfino et
al~\cite{DelfinoBarkemaCardy00} (p.533) explain that their results
are sensitive to the relative normalization of the order- and
disorder-operator form-factors, which could be the origin of some
troubles for $q=3$ and $4$ for the ratios $\Gamma_+/\Gamma_L$ and
$R_C$ only.

As a final argument in favor of our results, we may mention a work
of W. Janke and one of us (LNS) on the amplitude ratios in the
Baxter-Wu model (expected to belong to the 4-state Potts model
universality class), leading to the estimate
$\Gamma_+/\Gamma_-\simeq 6.9$ and $R_C^-\approx
0.005$~\cite{BBJS06}. This result, obtained from an analysis of MC
data shows a similar discrepancy with Delfino and Cardy's result and
suggests that further analysis might still be necessary.

\section{Acknowledgements}

Discussions with A. Zamolodchikov, V. Dotsenko, V. Plechko, W. Janke
and M. Henkel, and a correspondence with J. Cardy and J. Salas were
very helpful.

LNS is grateful to the Statistical Physics group of the University
Henri Poincar\'e Nancy~1 for the kind hospitality. Both LNS and PB
thank the Theoretical group of the University Milano--Bicocca for
hospitality and support. Financial support from the
Laboratoire Europ\'een Associ\'e
``Physique Th\'eorique et Mati\`ere Condens\'ee'',  a common research
program between the Landau Institute, the Ecole Normale
Sup\'erieure de Paris, Paris Sud University and the
Russian Foundation for Basic Research is
also gratefully acknowledged.

\newpage

%
%

\appendix

\section{Solution of non-linear RG equations and cancelation of logarithmic
corrections in effective-amplitude ratios}
\label{ap1}

For the 4-state Potts model, the non-linear RG equation for
the relevant thermal and magnetic fields $\phi$ and $h$, with the corresponding
RG eigenvalues $\yT $ and $\yH $, and
the marginal dilution field $\psi$ are given by
\begin{eqnarray}
    \frac{d\phi }{d \ln b}&=&(\yT +\yTpsi \psi )\phi
        ,\label{ap-eq2}\\
    \frac{dh }{d \ln b}&=&(\yH +\yHpsi \psi)h,
        \label{ap-eq3}\\
    \frac{d\psi }{d \ln b}&=&g(\psi)
        \label{ap-eq1}
\end{eqnarray}
where $b$ is the length rescaling factor and $l=\ln b$.  The
function $g(\psi)$ may be Taylor expanded, $g(\psi)=\ypsideux
\psi^2(1+\frac{\ypsitrois }{\ypsideux }\psi+\dots)$. Accounting for
marginality of the dilution field, there is no linear term. The
first term has been considered by Nauenberg and Scalapino, and later
by Cardy, Nauenberg and Scalapino. The second term was introduced by
Salas and Sokal. In this appendix, we slightly change the notations
of Salas and Sokal, keeping however the notation $y_{ij}$ for all
coupling coefficients between the scaling fields $i$ and $j$. These
parameters take the values  $\yTpsi =3/(4\pi)$, $\yHpsi =1/(16\pi)$,
$\ypsideux =1/\pi$ and $\ypsitrois =-1/(2\pi^2)$, while the relevant
scaling dimensions are $\yT =\nu^{-1}=3/2$ and $\yH =15/8$.

The fixed point is at $\phi=h=0$. Starting from initial conditions
$\phizero$, $\hzero$, the relevant fields grow exponentially with
$l$. The field $\phi$ is analytically related to the temperature,
so the temperature behavior follows from the renormalization flow
from $\phizero\sim|\tau|$ up to some $\phil=O(1)$ outside the
critical region. Notice also that the marginal field $\psil$
remains of order $O(\psizero)$ and $\psizero$ is negative,
$\psizero=O(-1)$. In zero magnetic field, under a change of length
scale, the singular part of the free energy density transforms
as \be
    f(\psizero,\phizero)=e^{-Dl}f(\psil,\phil),\label{ap-eq4}
\ee
where $D=2$ is the space dimension.
Solving \Eq{ap-eq2} leads to
$\ln(\phil/\phizero)=\yT l+\yTpsi \int\psi dl$ where the last integral is
obtained from \Eq{ap-eq1} rewritten as
$\int_0^l\psil dl=\frac 1{\ypsideux }\ln ({\psil}/{\psizero})
+\frac{1}{\ypsideux }\ln G(\psizero,\psil)$.
Note that $G(\psizero,\psil)$ takes the value 1 in Cardy, Nauenberg and
Scalapino and the value $\frac{\ypsideux +\ypsitrois \psizero}
{\ypsideux +\ypsitrois \psil}$ in Salas and Sokal.
Since this term appears  always in the same combination,
we write $z=\psizero/\psil$, $\bar z=\frac z{G(\psizero,\psil)}$
and in the same way we set $x=\phizero/\phil$.
One thus obtains
\be
    l=-\frac 1{\yT }\ln x+\frac{\yTpsi }{\yT \ypsideux }
    \ln \bar z.
    \label{ap-eq5}
\ee
At the critical temperature $\phi=0$, the $l-$dependence on the magnetic
field obeys a similar expression and one is led to the equality
\be
    l=-\nu 
    \ln x+\mu 
    \ln\bar z=-\nuH 
    \ln y+\muH 
    \ln\bar z,\label{eq-xy}
\ee
where $y=\hzero/\hl$ and for brevity we will denote
$\nu=1/\yT=\frac 23$, $\mu=\frac{\yTpsi }{\yT\ypsideux }=\frac 12$,
$\nuH=1/\yH=\frac{8}{15} $ and
$\muH=\frac{\yHpsi }{\yH\ypsideux }=\frac{1}{30}$.
We can thus deduce the following behavior for the free energy density
in zero magnetic field in
terms of the thermal and dilution fields, or at the critical
temperature in terms of magnetic and
dilution fields
\begin{eqnarray}
    f(\phizero,\psizero)&=& x^{D\nu}\ \!
    \bar z^{-D\mu}
    f(\phil,\psil)
    \label{ap-eq6}\\
    f(\hzero,\psizero)&=& y^{D\nuH}\ \!
    \bar z^{-D\muH}
    f(\hl,\psil)
    \label{ap-eq6bis}
\end{eqnarray}

The other thermodynamic properties follow by derivation with respect to
the scaling fields, e.g.
$E(\phizero,\psizero)=\frac \partial{\partial \phizero}f(\psizero,\phizero)$, or
\begin{eqnarray}
E(\phizero,\psizero)
    &=&e^{-Dl}\frac{\partial\phil}{\partial\phizero}E(\phil,\psil),
    \label{eq-dfnE}\\
C(\phizero,\psizero)
    &=&e^{-Dl}\left(\frac{\partial\phil}{\partial\phizero}\right)^2
    C(\phil,\psil),
    \label{eq-dfnC}\\
M(\phizero,\psizero)
    &=&e^{-Dl}\frac{\partial \hl}{\partial \hzero}M(\phil,\psil),
    \label{eq-dfnM}\\
\chi(\phizero,\psizero)
    &=&e^{-Dl}\left(\frac{\partial \hl}{\partial \hzero}\right)^2
    \chi(\phil,\psil).
    \label{eq-dfnchi}
\end{eqnarray}
The derivatives $\frac{\partial\phil}{\partial\phizero}$
and $\frac{\partial \hl}{\partial \hzero}$ will thus determine the scaling
behavior of all the thermodynamic quantities.
The first one is obvious, $\frac{\partial\phil}{\partial\phizero}=x^{-1}$ and
for the second one, $\frac{\partial \hl}{\partial \hzero}=y^{-1}$,
we express $y$ in terms of
$x$ and $z$ using eq.~(\ref{eq-xy})~\footnote{It follows
that $y=x^{\yH /\yT }\bar z^{\yHpsi /\ypsideux -
\yH \yTpsi /\yT \ypsideux }$.}.
Altogether, introducing the  notations
$\lambda=\yH /\yT=\frac 54 $ and
$\kappa=\yH \yTpsi /\yT \ypsideux -\yHpsi /\ypsideux =\frac 78$
one has
\begin{eqnarray}
f(\phizero,\psizero)    &=&x^{D\nu}\ \! \bar z^{-D\mu}f(\phil,\psil)\\
E(\phizero,\psizero)    &=&x^{D\nu-1}\ \! \bar z^{-D\mu}E(\phil,\psil)\\
C(\phizero,\psizero)    &=&x^{D\nu-2}\ \! \bar z^{-D\mu}C(\phil,\psil)\\
M(\phizero,\psizero)    &=&x^{D\nu-\lambda}\ \! \bar z^{-D\mu+\kappa}
    M(\phil,\psil)\label{eq-M96}\\
\chi(\phizero,\psizero) &=&x^{D\nu-2\lambda}\ \! \bar z^{-D\mu+2\kappa}
    \chi(\phil,\psil)
\end{eqnarray}
What appears extremely useful in these expressions is that when defining
appropriate effective ratios\footnote{i.e. effective ratios which
eventually tend
to universal limits when $\tau\to 0$}, the dependence on the quantity
$\bar z$ cancels, due to the scaling
relations among the critical exponents. This quantity $\bar z$
is precisely the only
one where the log terms are hidden, and thus
we may infer that not only the leading log terms, but all the log terms hidden
in the dependence on the marginal dilution field disappear in the conveniently
defined effective ratios.
For example in an effective ratio like
\be R_C(x)=x^{2}\frac{C(x,z)\chi(x,z)}{M^2(x,z)}\label{rat-app},\ee
all corrections to scaling coming from the variable $z$  disappear.

Now we proceed by iterations of $l=-\nu\ln x+\mu\ln\bar z$ and
$\bar z=z\frac{1+(\ypsitrois /\ypsideux )\psil}
{1+(\ypsitrois /\ypsideux )\psizero}$.
The asymptotic solution of \Eq{ap-eq1}
is\footnote{When $\ypsitrois=0$ the asymptotic solution of \Eq{ap-eq1}
is simply
$\frac \psizero{1-\psizero \ypsideux l}$.
Thus one can try the ansatz
$\psi=\frac \psizero{1-\psizero \ypsideux l}(1+X(l))$ with a small
correction $X(l)$ to solve asymptotically the full \Eq{ap-eq1}.
Keeping only terms of order $O(l^{-3})$ at most, we are led to
the following expression $X'(l)=-(1/l)X(l)+\ypsitrois/((\ypsideux)^2l^2)$
where we use
$X(l)=Y(l)/l$ to eventually obtain $Y(l)=(\ypsitrois/(\ypsideux)^2)\ln l$.
}
\be
    \frac\psil\psizero=\frac 1{1-\psizero \ypsideux l}
    \left(1+\frac{\ypsitrois }{(\ypsideux)^2}\frac{\ln l}l +O(1/l)\right).
\ee
We get for the variable $z$
\begin{eqnarray}
    z=\frac{
        1-\psizero \ypsideux l}
        {1+ \frac{\ypsitrois }{(\ypsideux)^2}
            \frac{\ln l}l
        }
    \simeq -\psizero \ypsideux \nu (-\ln|\tau|)
    \frac{
        1+\frac \mu\nu
        \frac{\ln(-\ln|\tau|)}
            {-\ln|\tau|}
    +O\left(\frac 1{-\ln|\tau|}\right)
    }
    {1+\frac{\ypsitrois }{(\ypsideux)^2\nu}
    \frac{\ln(-\ln|\tau|)}{-\ln|\tau|}
    +O\left(\frac 1{-\ln|\tau|}\right)
    }.
\end{eqnarray}
Similarly, one has the combination
\be
    \frac {1+(\ypsitrois /\ypsideux )\psil}
    {1+(\ypsitrois /\ypsideux )\psizero}
    \simeq
    \frac{1}{1+(\ypsitrois /\ypsideux )\psizero}
    \left(
    1-\frac{\ypsitrois }{(\ypsideux)^2\nu}\frac{1}{(-\ln|\tau|)}
    +O\left(\frac{1}{(-\ln|\tau|)^2}\right)\right)
\ee
and eventually one gets for the full correction-to-scaling variable
the {\em heavy} expression
\begin{eqnarray}
    \bar z&=&{\rm const}\times
    (-\ln|\tau|)
    \frac{
        1+\frac \mu\nu
        \frac{\ln(-\ln|\tau|)}
            {-\ln|\tau|}
    }
    {1+\frac{\ypsitrois }{(\ypsideux)^2\nu}
    \frac{\ln(-\ln|\tau|)}{-\ln|\tau|}
    }
    \times
    \left(
    1-\frac{\ypsitrois }{(\ypsideux)^2\nu}\frac{1}{(-\ln|\tau|)}
    \right)\times {F}(-\ln|\tau|)\nonumber \\
    &=&
    {\rm const}\times
    (-\ln|\tau|)
    \frac{
        1+\frac 34
        \frac{\ln(-\ln|\tau|)}
            {-\ln|\tau|}
    }
    {1-\frac 34
    \frac{\ln(-\ln|\tau|)}{-\ln|\tau|}
    }
    \times
    \left(
    1+\frac 34\frac{1}{(-\ln|\tau|)}
    \right)\times {\cal F}(-\ln|\tau|)
\end{eqnarray}
where ${\cal F}(-\ln|\tau|)$ is a function of the variable
$(-\ln|\tau|)$ only where also appears the non-universal constant
$\psizero$. Using \Eq{eq-M96}, we deduce the behavior of the
magnetization for example
\begin{eqnarray}
    M(-|\tau|)&=&B|\tau|^{1/12}(-\ln|\tau|)^{-1/8}
    \left[
    \left(1+\frac 34\frac{\ln(-\ln|\tau|)}{-\ln|\tau|}\right)
        \right.\nonumber\\
    &&\left.
    \ \qquad
    \left(1-\frac 34\frac{\ln(-\ln|\tau|)}{-\ln|\tau|}\right)^{-1}
    \left(1+\frac 34\frac{1}{-\ln|\tau|}\right)
    {\cal F}(-\ln|\tau|)
        \right]^{-1/8}.
\end{eqnarray}
Since all these log expressions are ``lazy functions'', it is unsafe
to expand such terms, e.g. $\left(1-\frac
34\frac{\ln(-\ln|\tau|)}{-\ln|\tau|}\right)^{-1} \simeq 1+\frac
34\frac{\ln(-\ln|\tau|)}{-\ln|\tau|}$, since the correction term is
not small enough in the accessible temperature range $|\tau|\simeq
0.05-0.10$. We can thus only extract an effective function ${\cal
F}_{eff}(-\ln|\tau|)$ which mimics the real one ${\cal
F}(-\ln|\tau|)$ in the convenient temperature range. This is done
through a plot of an effective-magnetization amplitude
\begin{eqnarray}
    B_{eff}(-|\tau|)&=&M(-|\tau|)|\tau|^{-1/12}(-\ln|\tau|)^{1/8}
    \left[
    \left(1+\frac 34\frac{\ln(-\ln|\tau|)}{-\ln|\tau|}\right)
        \right.\nonumber\\
    &&\left.
    \qquad\qquad\qquad
    \left(1-\frac 34\frac{\ln(-\ln|\tau|)}{-\ln|\tau|}\right)^{-1}
    \left(1+\frac 34\frac{1}{-\ln|\tau|}\right)
        \right]^{1/8}
\end{eqnarray}
which is found to
behave as
\be
    B_{eff}(-|\tau|)=B\left(1-\frac{C_1}{-\ln|\tau|}
-\frac{C_2\ln(-\ln|\tau|)}{(-\ln|\tau|)^2}\right)^{1/8}\label{eq-BeffApp}
\ee from which one deduces that the function ${\cal F}(-\ln|\tau|)$
takes the approximate expression \be {\cal F}(-\ln|\tau|)\simeq
\left(1+\frac{C_1}{-\ln|\tau|}
+\frac{C_2\ln(-\ln|\tau|)}{(-\ln|\tau|)^2}\right)^{-1}. \ee What is
remarkable is the stability of the fit to \Eq{eq-BeffApp}. We obtain
  (see table~\ref{table_parameters_M})
$C_1=-0.757$ and $C_2=-0.522$ which yields an amplitude
$B=1.1570(1)$.   It is also possible to try a simpler choice, fixing
$C_1=0$ and approximating the whole series by the $C_2-$term only.
We then find the value   $C_2'=-0.88$   and this
leads to a very close magnetization-amplitude
  $B=1.1559(2)$.

For the following, we group all the terms coming from the variable
$\bar z$ into a single function ${\cal
G}(-\ln|\tau|)=(-\ln|\tau|)\times{\cal E}(-\ln|\tau|)\times {\cal
F}(-\ln|\tau|)$ where \bey
    {\cal E}(-\ln|\tau|)&=&
    \left(1+\frac 34\frac{\ln(-\ln|\tau|)}{-\ln|\tau|}\right)
        \nonumber\\
    &&
    \qquad\times \left(1-\frac 34\frac{\ln(-\ln|\tau|)}{-\ln|\tau|}\right)^{-1}
    \left(1+\frac 34\frac{1}{-\ln|\tau|}\right)
        \label{ap-eq99}
\eey
in terms of which the singular parts of the
physical quantities take a very compact form,
\bey
    f(\tau)&=&F_\pm |\tau|^{4/3}{\cal G}^{-1}(-\ln|\tau|)
        \label{ap-eq100}\\
    M(-|\tau|)&=&B |\tau|^{1/12}{\cal G}^{-1/8}(-\ln|\tau|)
        \\
    \chi_+(\pm |\tau|)&=&\Gamma_+ |\tau|^{-7/6}
        {\cal G}^{3/4}(-\ln|\tau|)
        \\
    \chi_L(-|\tau|)&=&\Gamma_L |\tau|^{-7/6}
        {\cal G}^{3/4}(-\ln|\tau|)
        \\
    \chi_T(-|\tau|)&=&\Gamma_T |\tau|^{-7/6}
        {\cal G}^{3/4}(-\ln|\tau|)
        \\
    E_\pm(\pm|\tau|)&=&   \frac{A_\pm}{\alpha (\alpha-1)}
      |\tau|^{1/3}
        {\cal G}^{-1}(-\ln|\tau|)
        \\
    C_\pm(\pm|\tau|)&=&   \frac{A_\pm}{\alpha}   |\tau|^{-2/3}
        {\cal G}^{-1}(-\ln|\tau|)
        .
    \label{ap-eq101}
\eey The function ${\cal E}$ is known exactly while the function
${\cal F}$ needs to be fitted to the numerical data. In the same
range of values of the reduced temperature, the ``correction
function'' ${\cal F}(-\ln|\tau|)$ is now {\em fixed and the only
remaining freedom  is to include background terms and possibly
additive corrections to scaling} coming from irrelevant scaling
fields~\footnote{To introduce corrections to scaling, let us
consider the case of an irrelevant scaling field, let say $g$,
coupled to the temperature field through
$$\frac{d\phi}{dl}=\yT\phi+\yTpsi\phi\psi + \yTg\phi g\quad
\hbox{and}\quad \frac{dg}{dl}=\yg g$$ ($\Delta >0$ above plays the
r\^ole of $-\yg/\yT$, and is thus linked to the corresponding
negative RG eigenvalue $\yg$). Solving for $g$ gives $\gl=\gzero
e^{\yg l}$ (the irrelevant scaling field decays exponentially when
one approaches the fixed point). Solving for $\psi$ gives
$\psil=\frac{\psizero}{1-\psizero\ypsideux l}$,
 and for $\phi$,
$$l=\frac 1\yT\ln(\phil/\phizero)-\frac{\yTpsi}{\yT\ypsideux}
\ln(\psil/\psizero)
+
\frac{\yTg}{\yT\yg}\gzero(e^{\yg l}-1).$$
Iteration now leads to
$$l=-\frac 1\yT\ln|\tau|+\frac{\yTpsi}{\yT\ypsideux}\ln(-\ln|\tau|)
+\frac{\yTpsi}{\yT\ypsideux}\ln\frac{|\psizero|\ypsideux}{\yT}
+\frac{\yTg \gzero}{\yT|\yg|}(1-|\tau|^{|\yg|/\yT})$$
and thus
a free energy density including the additive correction term
$$f\simeq e^{-Dl}={\rm const}\times|\tau|^{D/\yT}
(-\ln|\tau|)^{-D\yTpsi/\yT\ypsideux}
\left(1+\frac{D\yTg \gzero}{\yT|\yg|}|\tau|^{|\yg|/\yT}\right).$$
In our case, the $\ln|\tau|$ terms are due to the first scaling field
(marginal) through the complicated variable $\bar z$
and other correction terms could be added, e.g.
$$f(\tau)=F_\pm |\tau|^{4/3}  (-\ln\tmod)^{-1}{\cal E}^{-1}(-\ln|\tau|)
{\cal F}^{-1}(-\ln|\tau|)
(1+D|\tau|^{\frac 23|\yg|}). $$ }. Among the
additive correction terms, we may have those coming from the thermal
sector $\Delta_{\phi_n}=-\nu \yTn$, where the RG eigenvalues are
$\yTn=D-\frac 12 n^2$. The first dimension $\yTun=\yT$ is the
temperature RG eigenvalue, the next one, $\yTdeux$, vanishes and is
responsible for the appearance of the logarithmic corrections, so
the first irrelevant correction to scaling in the thermal sector
comes from $\Delta_{\phi_3}=-\nu\yTtrois=5/3$. One can also imagine
a coupling of the magnetic sector to irrelevant scaling fields. The
magnetic scaling dimensions $x_{\sigma_n}$ lead to RG eigenvalues
$\yHn=D-x_{\sigma_n}$. The first dimension $\yHun=\yH$ is the
magnetic field RG eigenvalue. The second one is still relevant,
$\yHdeux=7/8$, and it could lead, if admissible by symmetry, to
corrections generically governed by the {\em difference} of relevant
eigenvalues $(\yHun-\yHdeux)/\yT=2/3$. The next contribution comes
from $\yHtrois=-9/8$ and leads to a Wegner correction-to-scaling
exponent~\cite{Wegner72} $\Delta_{h_3}=-\nu\yHtrois=3/4$. Eventually, spatial
inhomogeneities of primary fields (higher order derivatives) bring
the extra possibility of integer correction exponents $y_n=-n$ in
the conformal tower of the identity. The first one of these
irrelevant terms corresponds to a Wegner exponent $\Delta_1=-\nu
(-1)=2/3$ and it is always present. We may thus possibly include the
following corrections: $|\tau|^{2/3}$, $|\tau|^{3/4}$,
$|\tau|^{4/3}$, $|\tau|^{5/3}$, \dots, the first and third ones
being always present, while the other corrections depend on the
symmetry properties of the observables.


\end{document}